\documentclass{article}

\usepackage{microtype}
\usepackage[utf8]{inputenc}
\PassOptionsToPackage{authoryear,compress,round}{natbib}
\usepackage[final]{neurips_data_2022}

\usepackage{siunitx}

\usepackage[pdftex, 			% Paramétrage de la navigation
pagebackref = false, 			% Ajoute les liens vers les pages dans la bibliographie
bookmarks = true, 			% Signets
bookmarksnumbered = false, 	% Signets numérotés
hyperindex = true,			% Hypertexte dans l'index
linktocpage = true,			% Le lien porte sur le numéro, pas sur le titre, dans la toc
pdfpagemode = UseNone, 		% Full Screen ou None
pdfstartview = Fit, 			% La page prend toute la largeur
pdfpagelayout = OneColumn,% Vue par page
colorlinks = true, 			%false, % Liens en couleur
urlcolor = blue, 				% Couleur des liens externes
citecolor = blue,
pdfborder = {0 0 0}, 			% Style de bordure : ici, pas de bordure
breaklinks = true,
]{hyperref} 					% Utilisation de HyperTeX

\usepackage[export]{adjustbox} % for valign=t

\usepackage[utf8]{inputenc} % allow utf-8 input
\usepackage[T1]{fontenc}    % use 8-bit T1 fonts
\usepackage{url}            % simple URL typesetting
\usepackage{booktabs}       % professional-quality tables
\usepackage{amsfonts}       % blackboard math symbols
\usepackage{nicefrac}       % compact symbols for 1/2, etc.
\usepackage{xcolor}         % colors

\title{Open High-Resolution Satellite Imagery: The WorldStrat Dataset -- With Application to Super-Resolution}

\author{%
Julien Cornebise \\
University College London \\
\& Why How Ltd \\
\small{\texttt{j.cornebise@ucl.ac.uk}}
\And
Ivan Oršolić\\
Why How Ltd \\
\small{\texttt{ivanorsolic@gmail.com}}
\And 
Freddie Kalaitzis\\
University of Oxford \\
\small{\texttt{freddie.kalaitzis@cs.ox.ac.uk}}
}

\usepackage[colorinlistoftodos]{todonotes}
\providecommand{\tightlist}{%
  \setlength{\itemsep}{0pt}\setlength{\parskip}{0pt}}
  
\usepackage{longtable}
\usepackage{multirow, array, booktabs}

\usepackage{graphicx}
\usepackage{wrapfig}
\makeatletter
\def\maxwidth{\ifdim\Gin@nat@width>\linewidth\linewidth\else\Gin@nat@width\fi}
\def\maxheight{\ifdim\Gin@nat@height>\textheight\textheight\else\Gin@nat@height\fi}
\makeatother
\setkeys{Gin}{width=\maxwidth,height=\maxheight,keepaspectratio}
\makeatletter
\def\fps@figure{htbp}
\makeatother

\graphicspath{{./}{images/}}

\newcommand{\sqkm}{km² } % TODO: use sinunitx for proper rendering

\usepackage{tocloft}
\usepackage{minitoc}
\nostcrule

\begin{document}
\dosecttoc
%\faketableofcontents

\maketitle
\setcounter{secttocdepth}{2}

\providecommand{\sqkm}{km² } % TODO: use sinunitx for proper rendering
\providecommand{\ssqkm}{km²} % TODO: use sinunitx for proper rendering

\begin{abstract}%   <- trailing '%' for backward compatibility of .sty file
  Analyzing the planet at scale with satellite imagery and machine learning is a dream that has been constantly hindered by the cost of difficult-to-access highly-representative high-resolution imagery. To remediate this, we introduce here the \textbf{WorldStrat}ified dataset. The largest and most varied such publicly available dataset, at Airbus SPOT 6/7 satellites' high resolution of up to 1.5 m/pixel, empowered by European Space Agency's (ESA) Phi-Lab as part of the ESA-funded QueryPlanet project, we curate nearly 10,000 \sqkm of unique locations to ensure stratified representation of all types of land-use across the world: from agriculture to ice caps, from forests to multiple urbanization densities. We also enrich those with locations typically under-represented in ML datasets: sites of humanitarian interest, illegal mining sites, and settlements of persons at risk. We temporally-match each high-resolution image with multiple low-resolution images from the freely accessible lower-resolution Sentinel-2 (S2) satellites at 10 m/pixel. 
  We accompany this dataset with an open-source Python package to: rebuild or extend the WorldStrat dataset, train and infer baseline algorithms, and learn with abundant tutorials, all compatible with the popular eo-learn toolbox. 
  We hereby hope to foster broad-spectrum applications of ML to satellite imagery, and possibly develop from free public low-resolution Sentinel-2 imagery the same power of analysis allowed by costly private high-resolution imagery. We illustrate this specific point by training and releasing several highly compute-efficient baselines on the task of Multi-Frame Super-Resolution. 
  License-wise, the high-resolution Airbus imagery is CC-BY-NC, while the labels, Sentinel-2 imagery, and trained weights are under CC-BY, and the source code and pre-trained models under BSD, to allow for the widest use and dissemination.
The dataset is available at \url{https://zenodo.org/record/6810791}, the software package at \url{https://github.com/worldstrat/worldstrat}, and homepage at \url{https://worldstrat.github.io}.
\begin{figure}[!ht]
  \centering
  \includegraphics[height=8em]{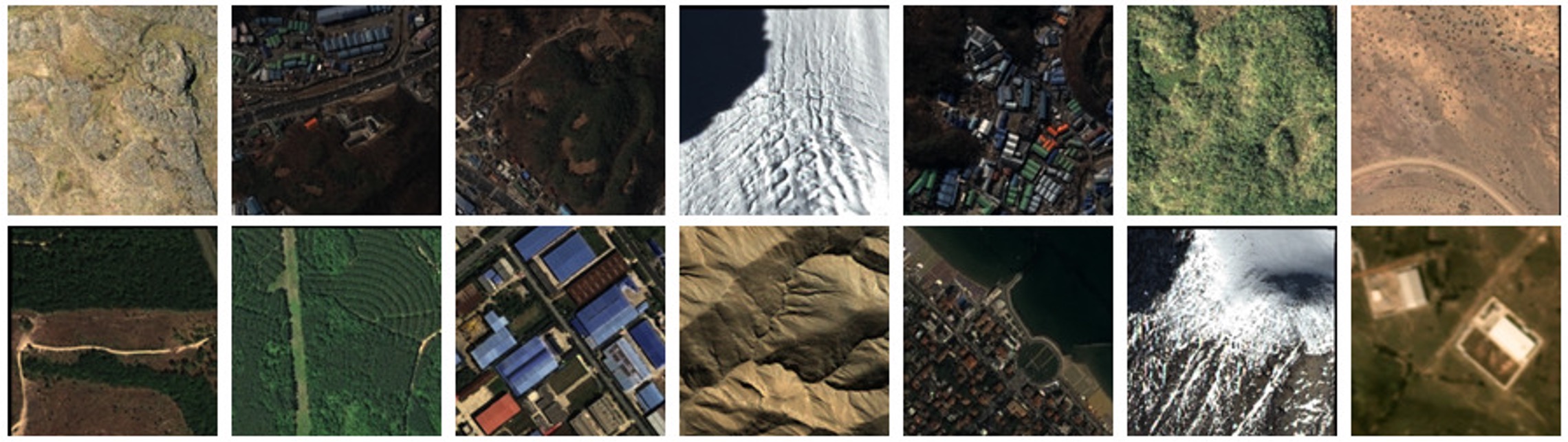}
  \caption{A glimpse at the variety of land uses covered by high-resolution imagery in the dataset.\label{fig:tile}}
\end{figure}

\end{abstract}
% To be rewritten, copied from a talk abstract Julien gave
%We present here our open-source package and models for Sentinel 2 multi-frame super-resolution, funded by the ESA QueryPlanet project. We include our trained neural networks with a focus on lightweight inference. By merging multiple revisits, we can trade temporal resolution for spatial resolution, allowing multiple use cases requiring higher details of static structures, at zero imagery cost, by everyone.
\ifdefined\neuripssolo
\vspace{-2em}
\else
\setcounter{tocdepth}{2}
\tableofcontents
\fi

% \begin{figure}[htbp]
%     \centering
%     \includegraphics[width=1\textwidthwidth]{WorldPOIs.png}
%     \caption{All locations represented in the dataset.\label{fig:partners_aois}}
% \end{figure} 
\section{Introduction} \label{sec:introduction}

\subsection{The Problem}

Computer vision and satellite imagery seem to be a match made in heaven. The idea to automatically process the growing amount of imagery collected has been lingering for decades in the remote sensing and the earth observation communities. The appeal of seeing the whole planet, and analyzing it at scale, is akin to few others. Many attempts have been made throughout the last thirty years. The ever-higher resolution of imagery available to civilians, and the last decade of improvement in Machine Learning and Computer Vision, have brought tools that could be brilliantly assistive in that regard. Some very visible scientific successes have been  published, such as \citep{jean_combining_2016}. Even prominent Deep Learning innovators cut their teeth early on as Master students on applications to satellite imagery, e.g.  \citet{mnih_learning_2010}. And prominent tools like Google Earth, which gives everyone access to high-resolution aerial imagery (although not all of it obtained by satellite), is a trigger for imagination. The dream runs deep.

However, the full potential of these two fields has been hindered by a combination of factors, in particular data access and the associated costs.

Broadening access to cutting edge technology is a hacker's delight: some satellite imagery can actually be received by any accessible for any amateur with an antenna up their roof -- back as an undergrad the first author built a cheap reception, storage, and processing station for low-resolution Meteosat Second Generation imagery \citep{beaudoin_meteosat_2005}, undercutting ten-fold the cost of commercial stations at the time. And while such refinements on reception stations are not feasible for high-accuracy digital imagery, since 2015, the European Sentinel-2 (S2) satellites have been providing medium-resolution imagery (10 m/pixel) for free every five days accross the world for anyone who knows how to access them. 

Yet, high-resolution imagery (1 m/pixel) or very-high-resolution (sub 1 m/pixel) are still out of easy reach. In a more recent work with Amnesty International to detect destroyed villages in conflict zones, \citep{cornebise_witnessing_2018}, we discovered that the cost to purchase a single very-high-resolution mosaic of the whole of Darfur, akin to Google Earth at maximum zoom, would cost 4 Million USD, even including a generous discount for charities. This makes it a tremendous challenge to even experiment with computer vision for high resolution.

Even setting cost aside, and assuming, as some hope, that the thunderous technological advances in launch technologies unlock a deluge a high-resolution imagery at a smaller price point, the key material for Machine Learning is still simply not there: carefully curated datasets to train on!  Accessing satellite imagery, even Sentinel 2, requires a certain amount of domain knowledge, more so than for natural images that can be sourced pretty much anywhere. The barrier to entry is real.

Indeed open high-resolution satellite imagery datasets are still quite rare, and the few who do exist tend to be either small, cover few unique locations, ad-hoc locations, or are designed for very specific uses. The SpaceNet challenge datasets \citep{van_etten_spacenet_2018} are widely used satellite imagery datasets. Their combined unique location area is close to WorldStrat with a bit more than 10 000 km², but it is mainly focused on urban structures and made for specific tasks like building or road detection, with varying data providers and resolutions, and with no paired multi-temporal low-resolution imagery.

We mention paired lower-resolution, because another hope to work around the lack of access lies in the field of Super-Resolution: being able to derive from (possibly multiple) free low-resolution satellite revisits the same insights that would be available from a single high-resolution satellite visit of the corresponding area.  While we want to build datasets that can be used for a whole breadth of applications, making them suitable for super-resolution brings a swath of extra benefits.

On that topic, the  ESA Kelvins PROBA-V dataset \citep{martens_super-resolution_2019} and the associated competition have been a boon for such multi-frame super-resolution, but is single-channel and most importantly is very low resolution at 300 m/pixel and 100 m/pixel. It is also not georeferenced or time-referenced. However, it allowed us to develop the high-performing HighRes-Net multi-frame super-resolution algorithm \citep{deudon_highres-net_2020}. There is a an abundant literature on multiframe super-resolution for natural images and videos, but quite not as much in satellite imagery -- which we will cover in Section~\ref{sec:superres}.

The very recent dataset by \citet{michel_sen2vens_2022} promises to enable single-image super-resolution. covers only 29 unique locations, on 806 unique \sqkm, at 5 meter/pixel. It offers nine revisits for each location, remarkably pairing one single low-resolution and one high-resolution for each visit. This is somewhat promising, but still lacks the breadth we hope for.

To the best of our knowledge, no dataset tries to be generic and cover systematically the whole type of land-use across the world. Even fewer are explicitly designed with the aim of transferring learning to high-availability lower-resolution data: low-resolutions from Sentinel 2 can be manually added, but at the price of  extra work and expertise.

\subsection{Our Contributions}

\begin{figure}[htbp]
  \centering
  \includegraphics[height=1\textwidth]{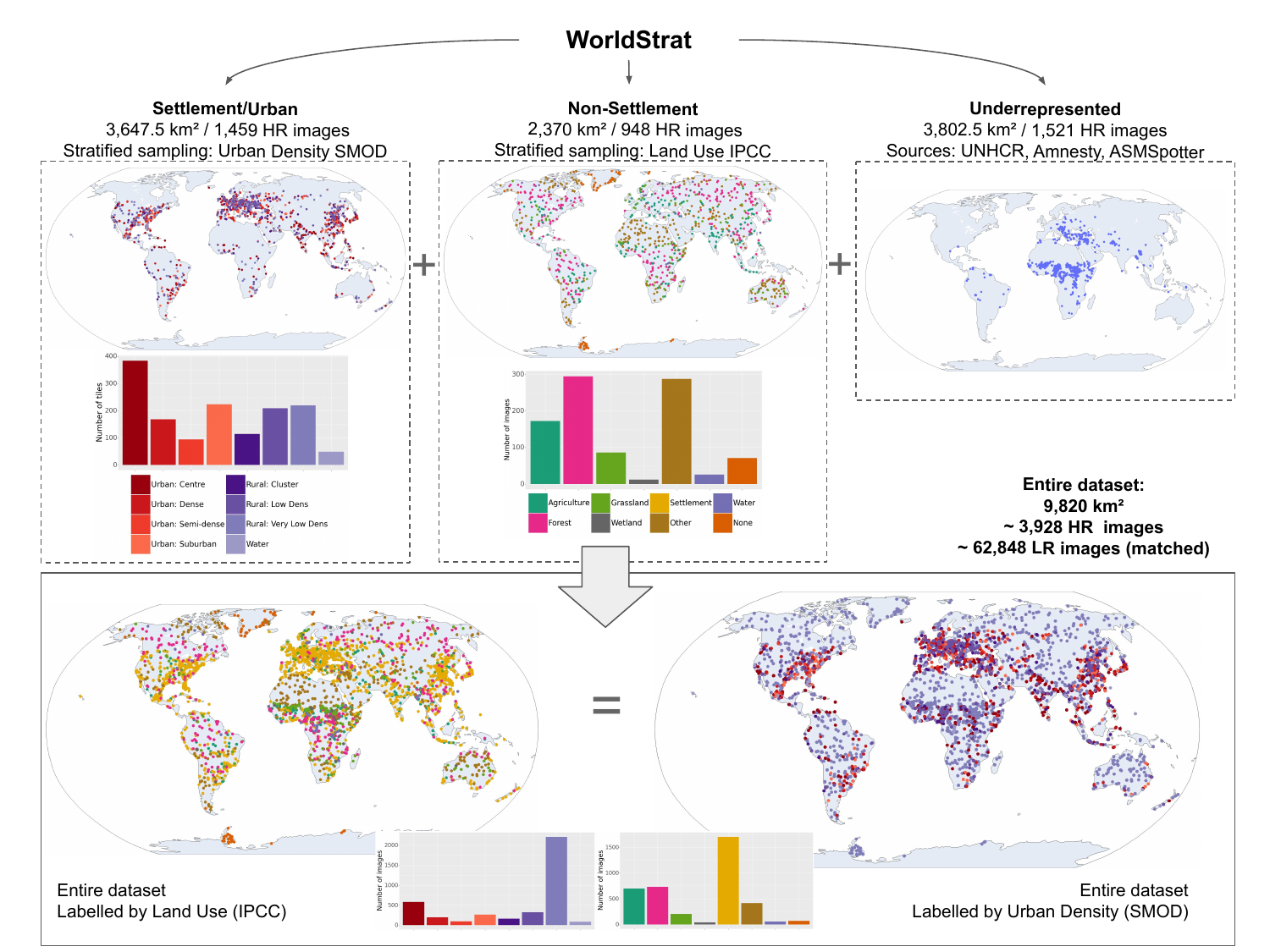}
  \caption{Summarizing the construction and classes of the WorldStrat dataset.\label{fig:summary_construction}}
\end{figure}

We aim to empower the development of machine learning for satellite imagery that can be used for a wide range of applications: from ecology and climate change monitoring, to urbanization, to sociology, to disaster preparation, to agriculture, etc.  Our focus while building this dataset and its accompanying benchmarks and software package has therefore been to: 
\begin{itemize}
  \tightlist 
  \item Maximise the representation of all possible features of interest, for the widest possible use-cases. 
  \item Have a decent worldwide representativity -- especially relevant in light of the need for Fairness Accountability and Transparency in ML, which includes the problem datasets often biased towards the Global North. 
  \item And provide a pipeline that allows easy reproducibility by others, and extension if extra budget becomes available.
\end{itemize}

Our resulting dataset, the \textbf{World Stratified Dataset} (or WorldStrat for short) \citep{cornebise_worldstrat_2022_zenodo} covers \textbf{$\sim$10,000 \ssqkm~, and 3504 distinct locations, specially curated for the highest diversity of possible uses}.   In particular, as visualised in Figure~\ref{fig:summary_construction}, we separated our image acquisition budget into three parts:
\begin{itemize}
\tightlist
  \item One part focusing on human settlements further stratified by population density: filtering the world for settlements according to the European Space Agency's (ESA) Climate Change Initiative (CCI) Land Cover product \citep{esa_land_v20} (which is licensed without restrictions under the CCI Data Policy v1.1), then sub-stratifying according to the Urban density class of the  Global Human Settlement Layer SMOD product \citep{florczyk_ghsl_2019} (which is licensed under CC-BY).
  \item A second part focused on non-settlement areas, using stratified sampling and class-rebalancing across the Land Cover Classification System labels of the ESA CCI Land Cover product, and more precisely its aggregation into classes provided by the International Panel on Climate Change (IPCC).
  \item Finally, a third part is focused on use cases typically under-served by usual datasets and not covered by the above. We sourced points of interests from the United Nations High Commissioner for Refugees (UNHCR) for populations of concerns (which is licensed under CC-BY-IGO), from Amnesty International for human rights sites of interest (with permission), and ESA artisanal and small-scale mining (ASM) ASMSpotter for illegal mining (with permission).
\end{itemize}

At each resulting location, along with the label, we provide the following imagery:
\begin{itemize}
  \tightlist
  \item One High-Resolution multispectral image from Airbus SPOT 6/7 (licensed a paid-for extension to ESA's Third Party Missions (TPM) license granting redistribution under CC-BY-NC), in RGB (6 m/pixel), Near Infrared (6 m/pixel), and Pan-chromatic channels (1.5m / pixel), at 1054x1054 pixels at the highest resolution.
  \item 16 Low-Resolution revisits from Copernicus Sentinel-2 (licensed without restrictions under Copernicus Sentinel Data Legal Notice and Service Information), temporally matched to the High-Resolution image -- within 5 days for the temporally closest. All 13 spectral bands are covered, at up to 10 m/pixel.
\end{itemize}

The rest of the article is structured as follows. In Section~\ref{sec:stratif}, we present how we have curated the parts of the world we cover, aka Areas Of Interest (AOIs), to offer maximum representativity of the world and of use cases.   In Section~\ref{sec:imagery}, we describe the characteristics of the paired imagery available at every AOI, both in low- and high-resolution, and how it can be easily extended.   In Section~\ref{sec:benchmarks}, to illustrate one possible use of this dataset, we establish baselines on multi-frame super-resolution tasks using several architectures, with an emphasis on compute efficiency. We also present the toolbox integrated with popular package eo-learn to use, reproduce, and extend, all our work. 
We conclude in Section~\ref{sec:discussion} by discussing ideas tried and discarded along the way, as well as possible extensions.  

\section{Curating Highly Representative Locations}
\label{sec:stratif}
The WorldStrat dataset covers almost 10,000 \sqkm. Each base AOI is 2.5 \sqkm, i.e. 1,581 meters per side, the minimum contiguous order size allowed by Airbus, provider of the high-resolutiom imagery. This maximizes the number of AOIs within the allotted budget.
% In order to maximise the variability of the features represented in the dataset, we collect AOIs at the smallest possible contiguous polygon size, which Airbus has set at 2.5 \sqkm, i.e. squares of 1,581 m side. This maximises the number of tiles we can obtain and the variety of features we can train on.

\subsection{Stratifying The World}
We use the first half of the dataset to attempt a systematic, stratified coverage of the world.
 The question becomes: how do we chose these locations to ensure a ''best'' application-agnostic dataset for super-resolution?
 
Sixty precent will be taken from the ``Settlement'' class from the ESA CCI LandCover Product, 
  which we then stratify according to the Global Human Settlement Layer (GHSL)
  Settlement Grid (SMOD) for different types of urban density, and with marginal distribution
  proportional to the cubic root of the actual distribution -\/- to
  keep the order of classes but diminish the overall imbalance. 
  
 Fourty  percent will be taken from all the other IPCC classes, i.e.
  non-settlement,
stratified according to (non-settlement) IPCC class, marginal
  distribution proportional to the cubic root of the actual
  distribution, and within each (non-settlement) ~IPCC class, again stratifying,
  according to the LCCS class (thinner vegetation typology), again with
  cubic root proportions.

\subsection{Why stratifying by land-use}
In an optimal sampling entirely focused on super-resolution, we would first find a latent-space feature-based clustering of satellite imagery, not unlike \citep{jean2019tile2vec}. Then, inspired by the classical statistical tools variance-minimizing design of experiments \citep{atkinson2007optimum}, we would sample each semantic cluster proportionally to how difficult it is to super-resolve. 

Quantifying this would, at best, be done via some form of active learning, with the problem of being dependent on the actual super-resolution algorithm used in the process, or at worst, require having first solved the super-resolution problem, neither of which is quite satisfying. 
 Besides, this would, again, be optimized solely for super-resolution: we aim for broader.

A proxy frequently used in statistical design of experiment is instead be to sample each semantic class proportionally to its variance. While this works well for numerical features, variance is not a clearly defined quantity for visual features. There would be an interesting methodological question to be investigated as to what a form of “semantic variance” would look like, but is quite a deep methodological topic and would take us into a rabbit hole out of scope of this project.

This ideal schema being quite ill-defined, we move on to a more pragmatic choice.

A pragmatic way to decide on an optimal dataset suitable for a broad range of applications is instead to ensure a representation of every type of land. This prepares equally for studies of urban space, of ice coverage, of limnology, etc.

We could sample POIs uniformly on POIs uniformly on the World Geodetic System 1984 (WSG84) ellipsoid, conditionally on being on land (by, say, rejection sampling), then filter the resulting points of interest (POIs) by SPOT availability. However this would not provide any guarantee as to how much each type of land use would be represented, nor in which proportion. The law of large numbers would guarantee a representation asymptotically proportional to the real world, but this does not provide any
fine-grained control, nor any representation of rare classes: random fluctuations would not be controlled for, which is problematic when the number of samples is ``far from asymptotic'' -- as in our case.

Instead, we can stratify by ``land use'', for some definition thereof, exploiting ample prior art on land use and land cover classifications.     

The ESA Climate Change Initiative Land Cover Products (ESA CCI LCP) \citep{esa_land_v20}. This worldwide land cover map is based on Food and Agriculture Organization of the United Nations (FAO) Land Cover Classification System (LCCS) hierarchical system \citep{di2005land}. Moreover, it nests it hierarchy within a coarser classification suggested by the International Panel for Climate Change \cite{esa_land_v20}[page 30.], reflecting the preoccupations of climate change studies.

\subsubsection{Sampling the Non-Urban World}

We sampled 2,000 \sqkm at 800 POIs and stratified them using the IPCC classification found in the ESA CCI Land Cover 2019 dataset. We then rebalanced the classes using cubic-root importance sampling, e.g. to under-represent on-purpose the surprisingly large planet-wide proportion of moss/lichen-covered land. The points were enriched to include ice-caps, noticeably absent from the original classification but of core importance both for climate analysis, geopolitical implications, and new logistics routes.

% TODO(julien): revisit two-columns vs one-column for LCCS classes after text is fixed
\begin{figure}[htbp]
  \centering
  \includegraphics[width=.48\textwidth]{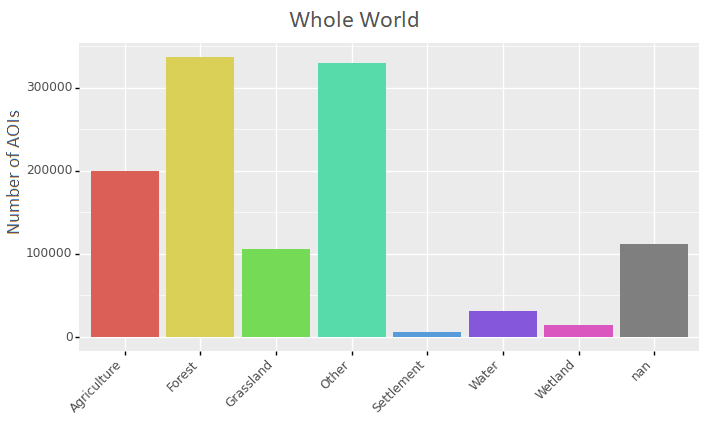}\hfill
  \includegraphics[width=.48\textwidth]{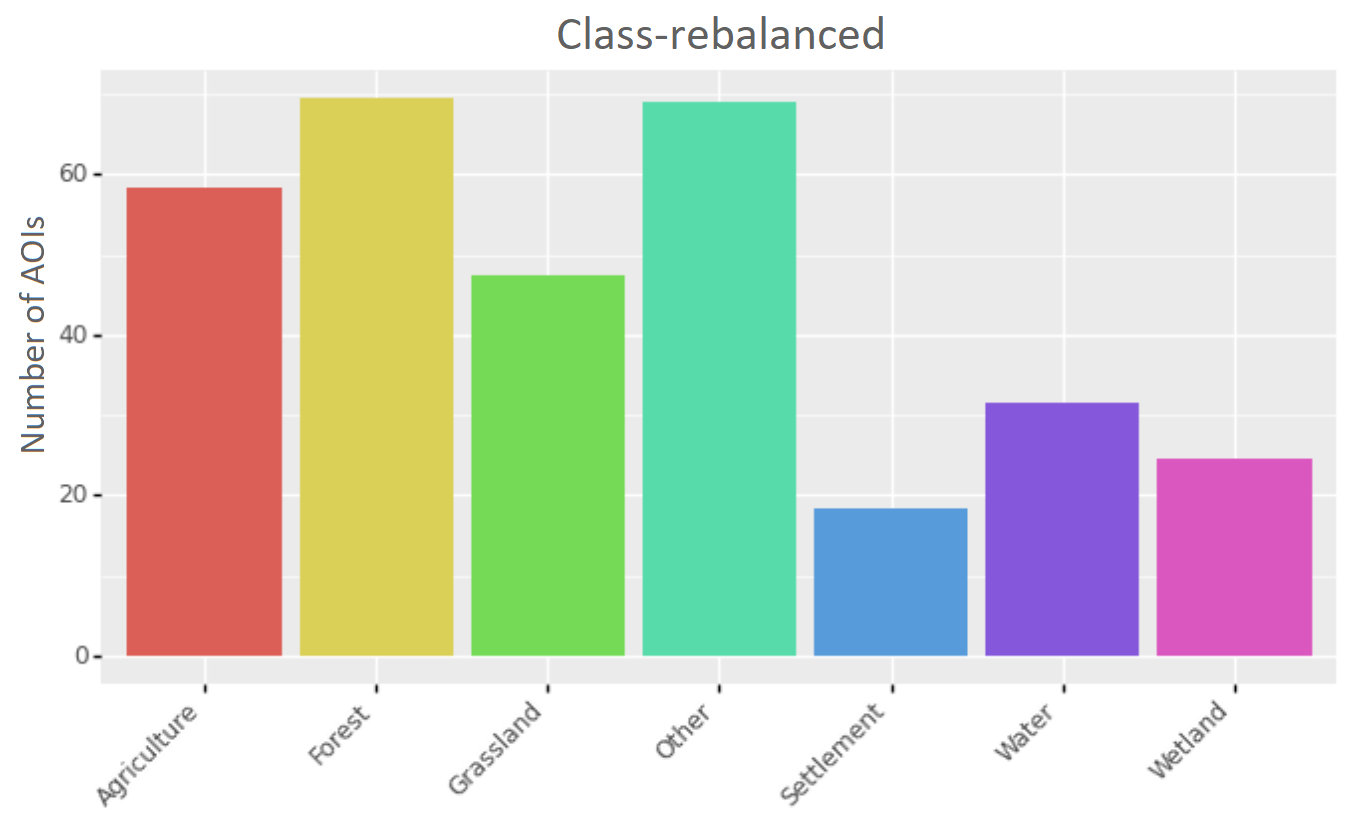}
  \caption{Distribution of the IPCC classes over the whole world (left), and the class-rebalanced sample using cubic root (right). \label{fig:lccs_ipcc_classes}}
\end{figure}

% \begin{figure}[htbp]
%   \centering
% \includegraphics[width=.75\textwidth]{images/lccs_ipcc_sqrt_classes.png}

%   \caption{}
%   \label{fig:lccs_ipcc_sqrt_classes}
% \end{figure}

The rationale for using importance sampling with the cubic root of
the population distribution as the proposal is the following: 

\vspace{-1em}
\begin{itemize}
\tightlist
\item We want to make sure we have a higher representation of rare classes,
  i.e. not being dominated by the classes that are already highly
  present naturally, because we wouldn't want our super-resolution to
  miss a rare object that would be out of place.
\item But we do still want to acknowledge that classes that are highly
  present naturally are possibly be super-resolved quite often, so we do
  not want to completely flatten the histogram by sampling uniformly.
\end{itemize}
Hence the cubic root as a compromise, which boosts rare classes but does keep the monotonous order relationship between the classes. The result is visible in Figure~\ref{fig:lccs_ipcc_classes}.

\subsubsection{Sampling the Urban World: Nested Stratification of Urban Density with GHSL SMOD}

We sampled 3,000 \sqkm at 1,200 POIs and filtered them by the "Urban" class found in the ESA CCI dataset. The points were then stratified by the density in Global Human Settlement Layer SMOD, which combines census and build-up data. The SMOD classes were also rebalanced by cubic-root importance sampling. We wanted to oversample the Settlement class, as it is quite crucial for a very wide range of applications, from geography to economics to demographics to aid planning to disaster recovery.

The Global Human Settlement Layer, an in particular its SMOD product \citep{florczyk_ghsl_2019}, focuses on density of building and population. A titanic work, it is not without  shortcomings, documented by~\citet{van_den_hoek_satellite-based_2021}, especially regarding informal settlements. Those would make it ill-suited for our needs, were it not complemented by the UNHCR-provided POIs.

Within the 1200 POIs classified as Settlement under the IPCC classification in ESA CCI LC, we stratify with respect to GHSL SMOD.  ESA CCI LC is of higher spatial resolution than SMOD.  But IPCC is of coarser semantic resolution than SMOD. SMOD has indeed multiple settlement classes, visible in Figure~\ref{fig:summary_construction}.

The relation in spatial resolution is inverse of the relation in land
use resolution. Therefore it makes extra sense to stratify IPCC settlement tiles (which
are of a smaller size than SMOD tiles) according to the type of (larger)
SMOD tile they fall into. This guarantees that we will have sampled
``IPCC settlements tiles'' from a varying type of surrounding urban
environment (as provided by SMOD).

\subsection{AOIs from Under-Represented Key Users}

The second half of the dataset is obtained by sourcing 3,895 sq km around 1,062 Points Of Interest (POIs) from specialists of use-cases ignored by most existing datasets. For the rarer type of POIs, we sample 9 actual images in a non-overlapping grid centered on the POI.

Non-governmental organizations (NGOs) and charities in Earth Observation (EO) interested about using recent Machine Learning articles are often met with the sobering answer ``Sorry, our models have never been trained on this type of landscape''. This is for example the case for informal settlements, refugee camps.  Most existing satellite imagery datasets, such as SpaceNet \citep{van_etten_spacenet_2018}, are often centred on urban features in large cities -- when they even cover beyond North America.

We also know that some of the highest benefits of this technique will go to social change actors who otherwise absolutely cannot afford high-resolution imagery.

For that reason,  we contacted high-potential users to inquire about the type of landscape they operate in. By obtaining examples of their past and ongoing points of interest (POIs) we ensure that our dataset will contain examples of the type of terrains in regions typically relevant to those users. 

\textbf{Amnesty International}, focusing on human rights violations and conflict areas: 495 \sqkm at 22 POIs, with 9 AOIs around each POIs.  Amnesty International’s geospatial specialist Micah Farfour, specialised in satellite monitoring of human rights abuse and environmental abuse, provided us with 22 POIs across the globe, ranging from barracks to prisons to mass grave sites. To make up for their small number, we ordered each of them on a 3x3 grid of minimum-size tiles, i.e. a total of 22.5 \sqkm per POI, with a total of 495 \sqkm.

\textbf{ASMSpotter}, focusing on illegal mining in remote areas. 900 \sqkm at 40 POIs, with 9 AOIs around each POIs, mostly in South America.  ASMSpotter's data specialist  Moritz Besser provided 40 POIs which we ordered similarly to Amnesty's, covering 900 \sqkm.

\textbf{United Nations High Commissioner for Refugees (UNHCR)}, focusing on informal human settlements in disaster and conflict areas. 2,500 \sqkm at 1,000 POIs, one AOI per POI. Oregon State University professor Jamon Van de Hoek, specialised in human settlements and in particular informal settlements -- see e.g. his latest publication \cite{van_den_hoek_satellite-based_2021} studying the Global Human Settlement Layer --
%The reaction from Pr. Van De Hoek to our request was revealing of the importance of our approach: he was overjoyed that we contacted him. He explained that all the authors of promising ML4EO methods he contacted sheepishly explained to him that their methods would not work on his problems, for lack of representation in the training dataset.  He 
kindly provided us with 3,000 POIs from the Persons of Concerns UNHCR Dataset~\citep{unhcr_2021}, which we downsampled to 1,000. 

Since certain parts of the world have a particularly high density of UNHCR-register location (e.g. in the Middle East), we avoided overlapping AOIs by ensuring all our AOIs are at least 10 km apart, using rejection sampling.

\section{Imagery}
\label{sec:imagery}

\vspace{-1em}
\subsection{High-Resolution: Single Visit SPOT 6/7}

\vspace{-1em}
Each AOI has a single visit of SPOT 6/7 high-resolution imagery, over five spectral bands.  The panchromatic (PAN) band has a resolution of 1.5 m/pixel, hence a 1,054x1,054 pixel image per 2.5 \sqkm AOI. The Red, Green, Blue, and Near Infrared bands (RGBNIR) are each at 6 m/pixel. 
The date of the visit has been picked at random between 2017 and 2019 amongst the visits whose whole-scene cloud-cover is lower than 5\%. Because our AOIs are much smaller than a full SPOT scene, it is not absolutely guaranteed that the actual image has precisely 5\% cloud -- it is likely to be entirely empty of clouds. This provides a good target image to reconstruct in the case of super-resolution. 
A glimpse of the imagery is show in Figure~\ref{fig:tile}.

\subsection{Low-Resolution: Multiple Revisits Sentinel 2}

\vspace{-1em}
For each high-resolution visit, we have 16 revisits. We picked revisits’ dates centered on SPOT visit. If more than 16 revisits per SPOT visit are needed, they are available via SentinelHub. The average time between revisits is 5 days.

The low-resolution revisits are provided in each of two product types: Level-1C (L1C) which provides top-of-atmosphere (TOA) reflectance images in cartographic geometry; Level-2A (L2A) which provides bottom-of-atmosphere (BOA) reflectance images derived from the associated Level-1C products by the European Space Agency. For each revisit, 13 bands are available in the L1C product type, and 12 bands in the L2A product type. The resolution ranges from from 10 m/pixel (for RGB) to 60 m/pixel.

 We chose to not filter the low resolution Sentinel-2 revisits by their cloud coverage. This is to try and ensure the training distribution on the low resolution is similar to the real world use cases, where the user will want to rebuild at a given place at a given time. Algorithms should learn to ignore clouds and be able to assemble a view from the cloudless parts of the cloudy revisits.

\subsection{Temporal Selection: Matching AOI, Low-Res, and High-Res Imagery}
%\todo{shorten this section and/or skip the graph}

Unlike Sentinel-2, SPOT is only available where and when it has been tasked. This raises two questions: How available and at which dates is SPOT high-resolution imagery of the POIs we have sampled? For each SPOT visit, how available is the Sentinel-2 imagery, within which time window around the date of the SPOT visit? 

\begin{figure}
    \centering
    \includegraphics[width=1\textwidth]{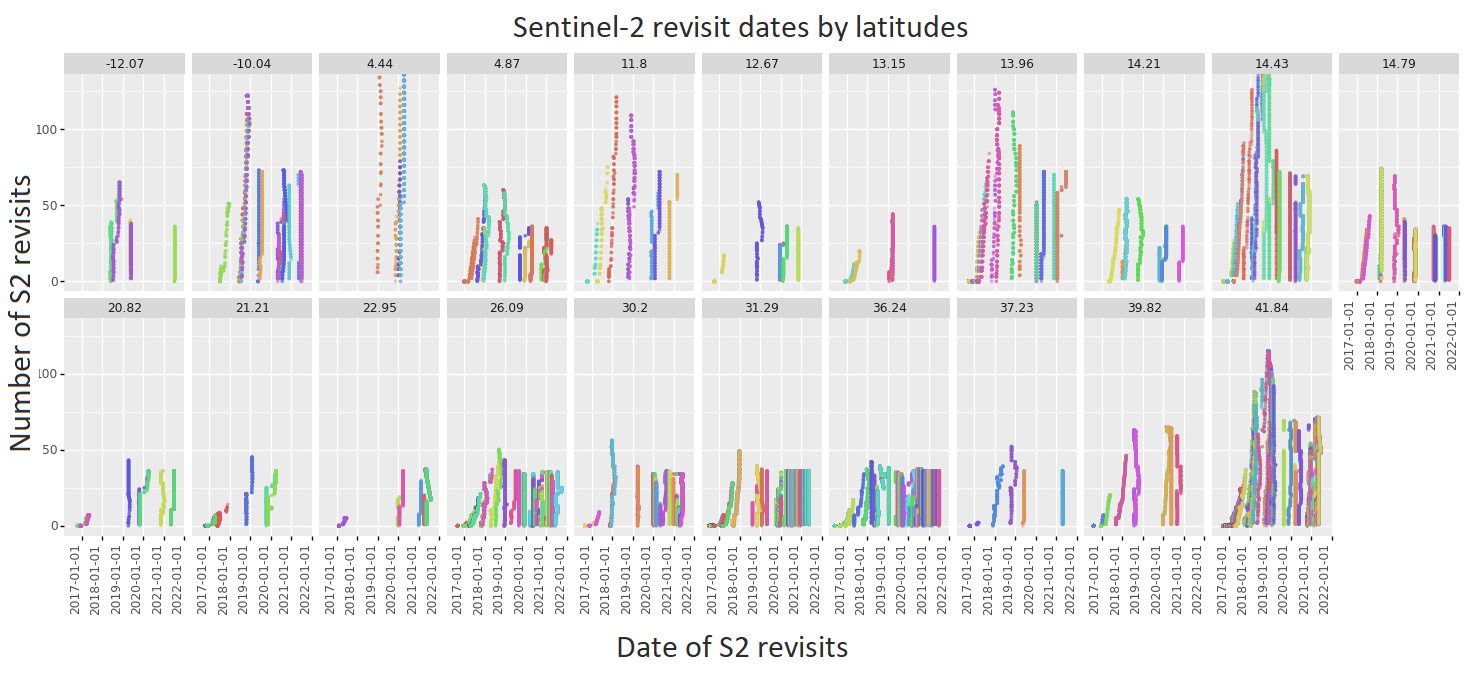}
    \caption{Number of Sentinel-2 visits for each SPOT visits on Amnesty International's AOIs. Each graph corresponds to an AOI and is labeled by its latitude at the top, to visualise the effect of location on revisits availability.}
    \label{fig:revisits}
\end{figure}

Figure~\ref{fig:revisits} illustrates the number of Sentinel-2 revisits (Y-axis) over each start and end date (up to +/- 6 months, coloured lines) around each SPOT visit (X axis at 0) of  each of the 22 Amnesty POIs (cell), in growing order of latitude (figured at the top of each cell). 

It shows that we could potentially try to pick POIs and SPOT visit times that maximise the number of Sentinel-2 imagery available within a fixed length time window, so as to have the richest training set. We do indeed observe some variation in that regard: some of these lines go much higher than others.

However, this would induce an implicit bias of a nature hard to interpret. We also observe that within a POI, the discrepancy between the number of S2 revisits, while clearly present, is reasonable, with multiple SPOT revisits offering similar S2 availability. 

Finally, biasing per Sentinel-2 imagery would be akin to biasing per Sentinel-2 cloud coverage: this would not be a fair representation of real-world use cases, and we would therefore be training our models for the wrong problem.

We therefore took the decision yet again to solve a harder problem than an optimally-curated dataset would make for, so as to be the closest to reality. To that effect, within a POI, we pick uniformly at random the SPOT visit to use as a reference.

Of course, one bias remains: we will not have imagery of POIs that have never been tasked by SPOT customers. While unfortunate, there is no way around it, short of using another high resolution product. We do have hope in two mitigating factors:
\begin{itemize}
\tightlist
  \item SPOT swath covers more than just the single POI, so we cover areas that are possibly more diverse than just the one precise point of interest to the SPOT customer.
  \item SPOT tasking means the POI exhibits features of activity interesting to at least the SPOT customer. SPOT customers might not have entirely the same interests as the users of our open-source package, but it is not unreasonable to assume that the features will be transferable. Therefore, this implicit sampling is actually a positive way to ensure interesting features.
\end{itemize}

\section{Putting it to Use: Baselines, Benchmarks, and Source Code}
\label{sec:using}

We believe this WorldStrat dataset can enable a broad range of applications. While we benchmark here multi-frame superresolution, closest to the authors' own expertise, this is by no means restrictive. More applications can be devised,  either with extra labelling, or using self-supervised representations on low and high resolutions, e.g. extending Tile2Vec \cite{jean2019tile2vec}, or even learning transfer tasks from one resolution to the other.

\subsection{Super-Resolution Benchmark}
\label{sec:superres}\label{sec:benchmarks}
We  illustrate the use of this dataset on the task of superresolution. While there has been considerable recent progress in multi-frame super-resolution (see \citet{freddie0, freddie1, freddie2, freddie3}), since this is not meant as an exhaustive benchmark article, we only focus on three architectures: 1) the single-image super-resolution architecture SRCNN \citep{dong2015image}; 2) our multi-frame extension of SRCNN (super-resolution convolutional neural network) by collating revisits as channels; and 3) a multi-spectral modification of the original HighResNet \citep{deudon_highres-net_2020} to handle multiple bands similarly to \citep{razzak_multi-spectral_2021}.

\begin{figure}
  \centering
  \includegraphics[width=.5\textwidth,valign=t]{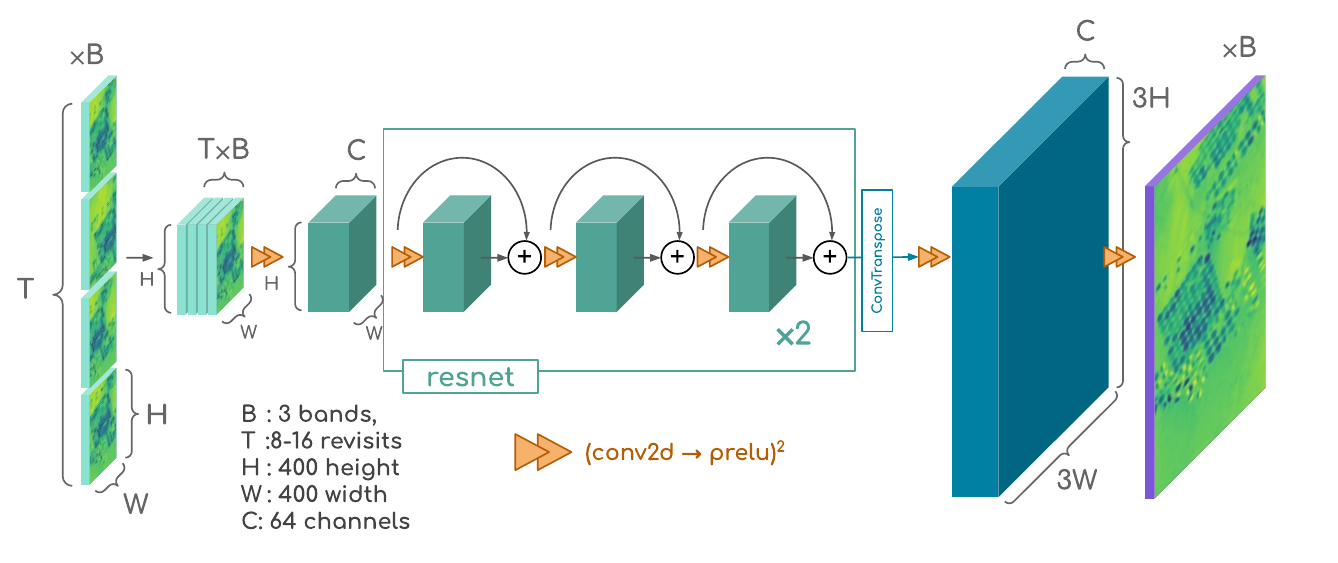}\hfill%
  \includegraphics[width=.5\textwidth,valign=t]{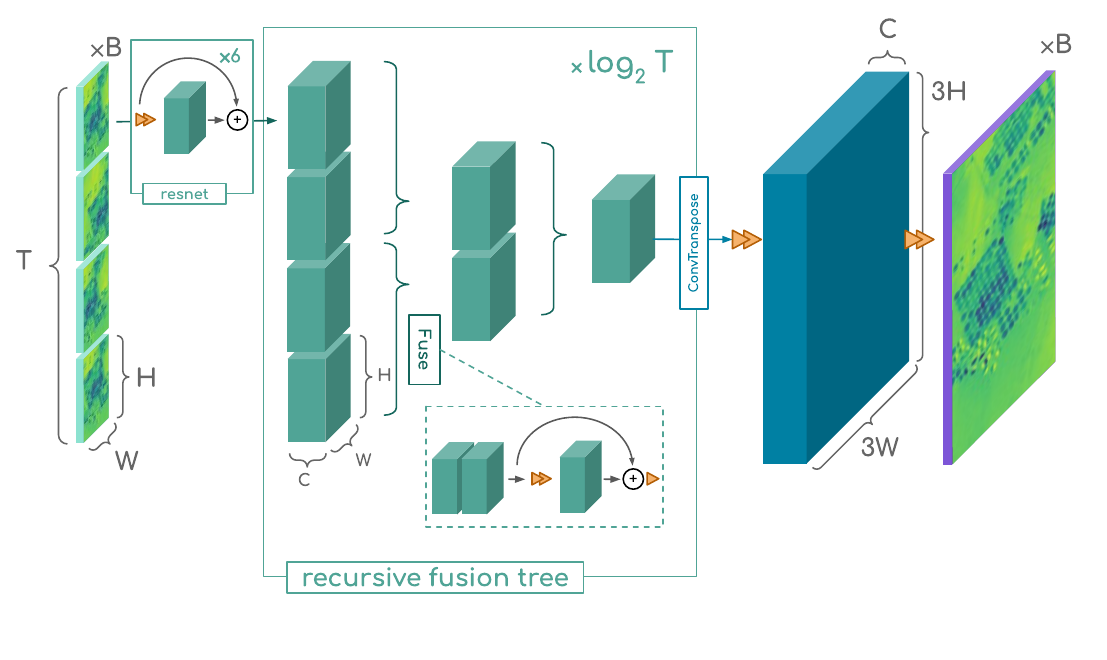}
  \caption{Multi-frame super-resolution architectures. Left: Multi-Frame SRCNN. Right: HighResNet. Single-frame SRCCN is achieved by using Multi-Frame SRCNN with a single revisit.\label{fig:architectures}}
\end{figure}

We accelerated HighResNet by replacing the learned ShiftNet by a simple cached alignment search. The two core architectures for (multi-frame) SRCNN and HighResNet are visualised in Figure~\ref{fig:architectures}. 
%\todo{cite the sources from Freddie, as we promised in the intro}

\begin{figure}[htbp]
  \centering
  \includegraphics[width=.5\textwidth]{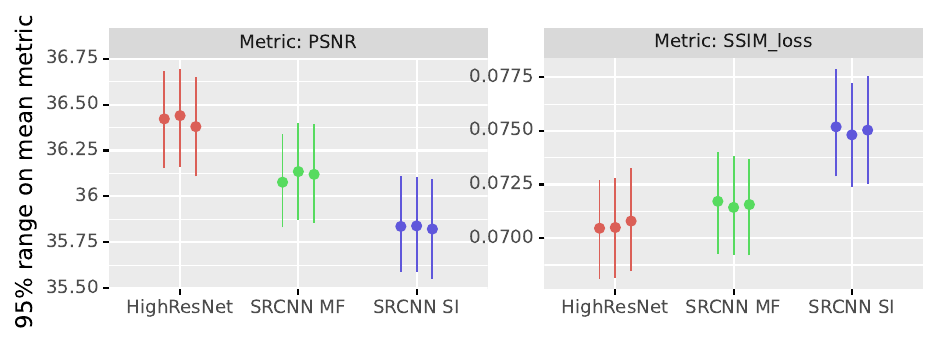}\hfill%
  \includegraphics[width=.5\textwidth]{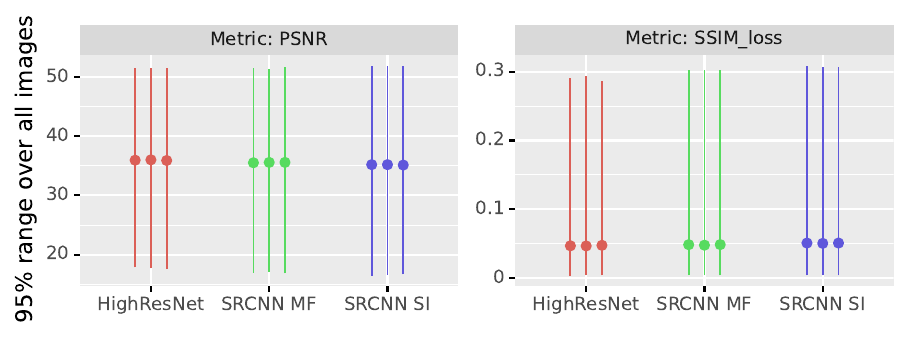}\hfill
  \caption{Comparing three architectures and two metrics over the validation set for three independent training runs. PSNR (Peak Signal-to-noise Ratio): higher is better. SSIM (Structural Similarity Index): lower is better. Left: comparing the 95\%  Confidence Intervals (CI) of the estimation of the mean of the metric. Right: comparing the 95\% range of the metric over the whole validation set. Note how, while the means seem significantly different, the variability across the distribution absolutely dwarfs any impact of the algorithms, pointing to the need for moving away from means-based benchmark, and to the need for better algorithms.\label{fig:ci_metrics}}
\end{figure}

All implementations are our own, and the code as well as the trained models are released as part of our eo-learn plugin, described in the following section. Figure~\ref{fig:ci_metrics} shows the results. 

We use the metrics established by~\cite{martens_super-resolution_2019}. Those consist of a weighted combination of mean average error (MAE), mean squared error (MSE), peak signal-to-noise ratio (PSNR) and structural similarity index (SSIM).

For each of the three architectures, we trained three models using three varying fixed random seeds, to account for results variability within an architecture. The same three seeds were used for every architecture, and a single fixed random seed was used for all data loading and processing operations, to ensure comparability between the architectures.

All training runs were done over the entire dataset, using 12-bit radiometry and 8 low-resolution revisits, using the published train, validation and test splits. Class stratification is ensured within the train, validation and test splits.  We used the Adam optimizer with a Cosine Annealing Warm Restart scheduler \citep{loshchilov_sgdr_2017}. We also provide Stochastic Gradient Descent (SGD) as an alternative optimizer. 

The benchmarks are reproducible using the provided open-source toolbox, and we provide all the hyperparameters in the Supplementary Material.

Rather than providing just the mean of each metric over the validation set (possibly with 95\% confidence intervals), we decide here to follow a harder but, we believe, much more rigorous reporting: we report the \emph{95\% range of the metric over the whole validation set}, i.e. the 2.5-percentile and the 97.5-percentiles over all data points in the validation sample. This provides for a much more rigorous comparison of algorithms, by showing the full context of the metric over the whole distribution of inputs. The reader can thus better see whether a few extra points on a metric make a strong difference or are just minor compared to within-distribution variability. 

As a side note, ablation studies (not shown here) have confirmed that the three above architectures enjoy a clear improvement when going from 4 revisits to 8, but only minor diminishing returns when increasing to 16. Another ablation study on the size of the training set shows there is still room for improvement were extra budget WorldStrat to be extended.

\subsection{Toolbox and eo-learn Plugin}
In addition to the dataset, we open-source a Python package designed for ease of use. It integrates with the widely eo-learn Python package as one in eo-learn pipeline, and provides 
abundant tutorial notebooks.  We made this dataset to be reproducible and extensible, and thus include 
notebooks covering data collection, training, and inference of all baselines, with standardized interfaces in Pytorch Lightning. We provide the ability to  sample new training data as needed.
 Particular care was paid to High-Efficiency Training, to make this world accessible with modest computing budget: our variant of HighResNet \textbf{trains in  30 minutes on a single V100 GPU}, thanks to our fast caching mechanism for 95\% average GPU usage.

\section{Discussion}
\label{sec:discussion}

\vspace{-1em}
\textbf{Potential social impact:}
A whole coverage of the potential social impact of computer vision and remote sensing could cover a whole book. However, in our specific case, the only sensitive data are the UNHCR locations, but these have already been released by UNHCR themselves. 

\textbf{Known limitations and future work:}
Although we want a broad range of applications to be possible with our trained models, we know that the frequency of revisits constrains us to structures that change at a slower pace than Sentinel-2 revisits -- one revisit every five days. This puts more emphasis on permanent structures, such as buildings or a form of land occupation, and rules out higher temporal variability use cases, such as immediate state of crops.

We did not enrich the dataset with rivers/harbours and liminal coastal space. This could be done by finding maps of water courses, and selecting locations at these intersections. 

Stratification could be endlessly refined. For example we considered but discarded stratifying by use of the  Local Climate Zones (LCZ) database of the World Urban Database
\citep{stewart_local_2012}. LCZ tries to identify the type of buildings on a
local scale, and is available for Europe, The Americas, and contributed zones
\citep{demuzere_lcz_2021}. However, the coverage of the contributed zones is not as wide as the GHSL.

The UNHCR Persons of Concerns dataset is not guaranteed to be extremely precise: its POIs are more likely ``in the neighbourhood'' of the actual settlements. We mitigate this by covering a 2.5 \sqkm area, and by accounting for the fact that temporary settlements are often similar to the surrounding settlements, which are therefore representative.

% Acknowledgements should go at the end, before appendices and references

\acksection{Project empowered by the ESA Phi Lab (\url{https://philab.phi.esa.int}) as part of the ESA-funded QueryPlanet project 4000124792/18/I-BG CCN3.
%\addstarredsection{Acknowledgements and Disclosure of Funding}
%\ifdefined\decrementstc
%\decrementstc
%\fi

We are very grateful to Pierre-Philippe Matthieu (ESA Phi-Lab) and Nicolas Longepe (ESA Phi-Lab) for their support and belief in this project. To Grega Milcinski (Sinergise) and SentinelHub for taking us on board QueryPlanet. To Jamon Van Den Hoek (Oregon State University) for his expertise on GHSL and providing the UNHCR POIs dataset. To  Micah Farfour (Amnesty International) for POIs of humanitarian interest. To Moritz Besser (dida Datenschmiede GmbH) for their ASMSpotter data. And Peggy Fischer, Bryan Keary, and Montserrat Del Riego (ESA TPM) for their support in making the imagery available publicly.}

\addcontentsline{toc}{section}{References}
\bibliographystyle{abbrvnat}
\bibliography{zotero}

\appendix
\section{Downloading the Dataset and the Software Package}
%\secttoc
%\subsection{test2}
The dataset, along with its machine-readable metadata, is hosted on CERN-backed Zenodo data repository: \url{https://zenodo.org/record/6810791} \citep{cornebise_worldstrat_2022_zenodo}. Its long-term maintenance is discussed in the Datasheet.

The software package is available on GitHub at  \url{https://github.com/worldstrat/worldstrat}. This includes reproducible code for the Benchmarks of Section
\ifdefined\datasheetsolo
4 of \citep{cornebise_neurips_2022},
\else
\ref{sec:benchmarks},
\fi
following the ML Reproducibility Checklist \citep{pineau2020checklistv2, pineau2021improving}. 

The project also has its own website available at \url{ https://worldstrat.github.io/}, containing links to the dataset and software package download and information on how to cite.

The authors hereby state that they bear all responsibility in case of violation of rights, etc., and confirm that the data license is as follows: 
\begin{itemize}
    \tightlist
    \item The low-resolution imagery, labels, metadata, and pretrained models are released under Creative Commons with Attribution 4.0 International (CC BY 4.0)\footnote{\url{https://creativecommons.org/licenses/by/4.0/}};
    \item The high-resolution imagery from Airbus is distributed under Creative Commons Attribution-NonCommercial 4.0 International (CC BY-NC 4.0)\footnote{\url{https://creativecommons.org/licenses/by-nc/4.0/}};
    \item The code is distributed under BSD license.
\end{itemize}

\clearpage

\section{Datasheet}

%\subsection{Test}
% Color picked from the Datasets for Datasheets paper
\definecolor{darkblue}{RGB}{46,25, 110}

\newcommand{\dssectionheader}[1]{%
\phantomsection
   \noindent\framebox[\columnwidth]{%
      {\fontfamily{phv}\selectfont \textbf{\textcolor{darkblue}{#1}}}
   }
  \addcontentsline{toc}{subsection}{#1}
}

\newcommand{\dsquestion}[1]{%
\medskip
\phantomsection
    {\noindent \fontfamily{phv}\selectfont \textcolor{darkblue}{\textbf{#1}}}
  \addcontentsline{toc}{subsubsection}{#1}
\smallskip
    }

\newcommand{\dsquestionex}[2]{%
\medskip
\phantomsection
    {\noindent \fontfamily{phv}\selectfont \textcolor{darkblue}{\textbf{#1} #2}}
  \addcontentsline{toc}{subsubsection}{#1}
    \smallskip
}
\providecommand{\sqkm}{km² } % TODO: use sinunitx for proper rendering
\providecommand{\ssqkm}{km²} % short sqkm, without space at the end (if used everywhere, doesn't allow a space to be added after the command for some reason).

This Datasheet for Dataset follows the template from \cite{gebru2021datasheets}.

%\ifdefined\datasheetsolo
%It is a companion to the article \cite{cornebise_neurips_2022} describing the dataset.
%\input{where_to_download}
%\fi

\ifdefined\datasheetsolo
%% FULL TOC for independent datasheet
%\tableofcontents
%\listoffigures 
%\listoftables 
%
%\clearpage
\else
\vspace{1em}
% MINITOC for datasheet embedded in another document
\renewcommand{\stctitle}{}
\setcounter{secttocdepth}{3}

\renewcommand{\cftsubsecfont}{\bfseries}
% TODO: find out *why* the TOC does not appear!
\adjuststc[-4]
\secttoc

\fi

\dssectionheader{Changelog: Versions of the Dataset and Datasheet}
\begin{center}
\begin{tabular}{|l|l|p{8cm}|}
    \hline
    \textbf{Version} & \textbf{Date} & \textbf{Changes} \\
    \hline
    1.0 & 2022-07-13 & Initial dataset and datasheet release.\\
    \hline
    1.1 & 2025-05-20 & 
    Add this Changelog.\newline
    Add missing file for AOI \texttt{ASMSpotter-1-1-1}.\newline
    Replace the recommended train/test/val split:\newline
    - fix data leakage between sets (11 AOIs overlap)\newline
    - incorporate 38 AOIs present in the dataset but absent from the split,\newline
    - stratify the split across joint classes (source, IPCC, SMOD),\newline
    - and update the split description in the datasheet.\newline
    Add to the metadata the "joint class" used for data split.\newline
    Fix missing header of the first column of the metadata CSV file.\newline
    Add link to Github issues for erratum.\\
    \hline
\end{tabular}
\end{center}

%%%%%%%%%%%%%%%%%%%%%%%%%%%%%%%%%%%%%%%%%%

\dssectionheader{Motivation}

\dsquestionex{For what purpose was the dataset created?}{Was there a specific task in mind? Was there a specific gap that needed to be filled? Please provide a description.}

Analyzing the planet at scale with satellite imagery and machine learning is a dream that has been constantly hindered by the cost of difficult-to-access highly-representative high-resolution imagery. We introduced this dataset to remediate this. 

The aim was to create, with a reasonable budget, the largest and most varied such publicly available dataset. We wanted to provide as broad and application-agnostic a representation of the physical features of the world as possible, by curating nearly 10,000 \sqkm  of unique locations to ensure stratified representation of all types of land-use across the world: from agriculture to ice caps, from forests to multiple urbanization densities. We also enrich those with locations typically under-represented in ML datasets: sites of humanitarian interest, illegal mining sites, and settlements of persons at risk. 

One particular set of tools that we aim to enable this dataset is the broad creation of multi-frame super-resolution algorithms, to amplify the use of the freely accessible but low-resolution satellite imagery from the European-funded Sentinel 2 constellation.  We achieve this by pairing high-resolution from Airbus SPOT 6/7 satellites with temporally-matched low-resolution imagery from Sentinel 2. 

To further make machine learning on satellite imagery accessible, we accompany this dataset with an open-source Python package to: rebuild or extend the WorldStrat dataset, train and infer baseline algorithms, and learn with abundant tutorials, all compatible with the popular EO-learn toolbox. Our code for deep learning algorithms provided as baseline on this dataset trains in 60 minutes on a single V100 GPU.

We therefore hope to foster broad-spectrum applications of ML to satellite imagery, and possibly develop from free public low-resolution Sentinel2 imagery the same power of analysis allowed by costly private high-resolution imagery. We illustrate this specific point by training and releasing several highly compute-efficient baselines on the task of Multi-Frame Super-Resolution. 
   
See 
\ifdefined\datasheetsolo
The Introduction in the main body of the article
\else
Section~\ref{sec:introduction}
\fi
for a longer perspective -- we do not copy it here to avoid redundancy.

\dsquestion{Who created this dataset (e.g., which team, research group) and on behalf of which entity (e.g., company, institution, organization)?}

This dataset was created by:
\begin{itemize}
   \tightlist
   \item Julien Cornebise, Ph.D., Honorary Associate Professor at University College London, and CEO of Why How Ltd, his sole-owner scientific consulting company.
   \item Ivan Oršolić, independent contractor, working for Why How Ltd for this project.
   \item Freddie Kalaitzis, Ph.D., Senior Research Fellow at Oxford University, working for Why How Ltd for this project.
\end{itemize}
We were empowered by the European Space Agency's Phi-Lab, \url{https://philab.phi.esa.int/}.

\dsquestionex{Who funded the creation of the dataset?}{If there is an associated grant, please provide the name of the grantor and the grant name and number.}

   The creation of this dataset was funded by the European Space Agency (ESA), as part of the "QueryPlanet" project 4000124792/18/I-BG CCN3, with ESA Phi-Lab championing this project. ESA Third Party Mission (TPM) funded the  license extension from Airbus required for distribution of the SPOT (high-resolution) imagery. Sinergise Ltd contributed in kind by giving free access to a SentinelHub account to facilitate access to the Sentinel Imagery. Julien Cornebise contributed in kind by volunteering his time, and funded part of the costs via his company Why How Ltd.

\dsquestion{Any other comments?}

Some of the locations listed in this dataset were kindly provided by:
\begin{itemize}
   \tightlist
   \item Jamon Van De Hoek, Ph.D., Associate Professor at Oregon State University who indicated the UNHCR Persons of Concerns dataset.
   \item Micah Farfour, Special Advisor in Remote Sensing at Amnesty International who provided 22 locations of human rights interest.
   \item Moritz Besser, Machine Learning Consultant at dida Machine Learning who provided 40 locations of artisanal mining.
\end{itemize}
Providing locations does not engage their responsibility or that of their employers.

The creation of this dataset transforms, or includes data from pre-existing datasets, in particular:
   \begin{itemize}
      \tightlist
      \item Randomly samples a subset of locations from UNHCR People of Concerns \citep{unhcr_2021}. 
      \item Filters world-wide randomly sampled locations using data from ESA CCI LC \citep{esa_land_v20}.
      \item Filters world-wide randomly sampled locations using data from GHSL SMOD \citep{florczyk_ghsl_2019}.
      \item The high-resolution imagery included in this dataset comes from Airbus as part of their SPOT 6/7 product \citep{spot_user_guide}. 
      \item The low-resolution imagery included in this dataset comes from the European Space Agency as part of the Copernicus Sentinel2 product \citep{drusch2012sentinel}.
   \end{itemize}

%%%%%%%%%%%%%%%%%%%%%%%%%%%%%%%%%%%%%%%%%%
\bigskip
\dssectionheader{Composition}

\dsquestionex{What do the instances that comprise the dataset represent (e.g., documents, photos, people, countries)?}{ Are there multiple types of instances (e.g., movies, users, and ratings; people and interactions between them; nodes and edges)? Please provide a description.}

Each instance represents a patch of land on Earth of 2.5 \ssqkm, i.e. 1581 m per side.

\dsquestion{How many instances are there in total (of each type, if appropriate)?}

There are 3,449 instances.

There are two types with regard to size: 2.5\sqkm and 22.5\sqkm instances:
\begin{itemize}
   \tightlist
    \item There are 3,388 $\times$ 2.5\sqkm instances.
    \item There are 61 $\times$ 22.5\sqkm instances.
    \item Their combined total is 9,820.57\ssqkm.
\end{itemize}
The 22.5\sqkm instances can be split into a grid of 3-by-3 2.5\sqkm instances, which brings the total of 2.5\sqkm instances to 3,937.

There are four types with regards to their location source:
\begin{itemize}
   \tightlist
    \item 22 $\times$ 22.5\sqkm Amnesty instances or 198 $\times$ 2.5\sqkm instances.
    \item 39 $\times$ 22.5\sqkm ASMSpotter instances or 351 $\times$ 2.5\sqkm instances.
    \item 981 $\times$ 2.5\sqkm UNHCR instances.
    \item 2,407 $\times$ 2.5\sqkm randomly sampled/stratified instances.
\end{itemize}

\dsquestionex{Does the dataset contain all possible instances or is it a sample (not necessarily random) of instances from a larger set?}{ If the dataset is a sample, then what is the larger set? Is the sample representative of the larger set (e.g., geographic coverage)? If so, please describe how this representativeness was validated/verified. If it is not representative of the larger set, please describe why not (e.g., to cover a more diverse range of instances, because instances were withheld or unavailable).}

The population of all possible instances would be the full surface of the Earth, at all possible times -- so this dataset is very much a sample.

\begin{figure}[htbp]
  \centering
  \includegraphics[height=1\textwidth]{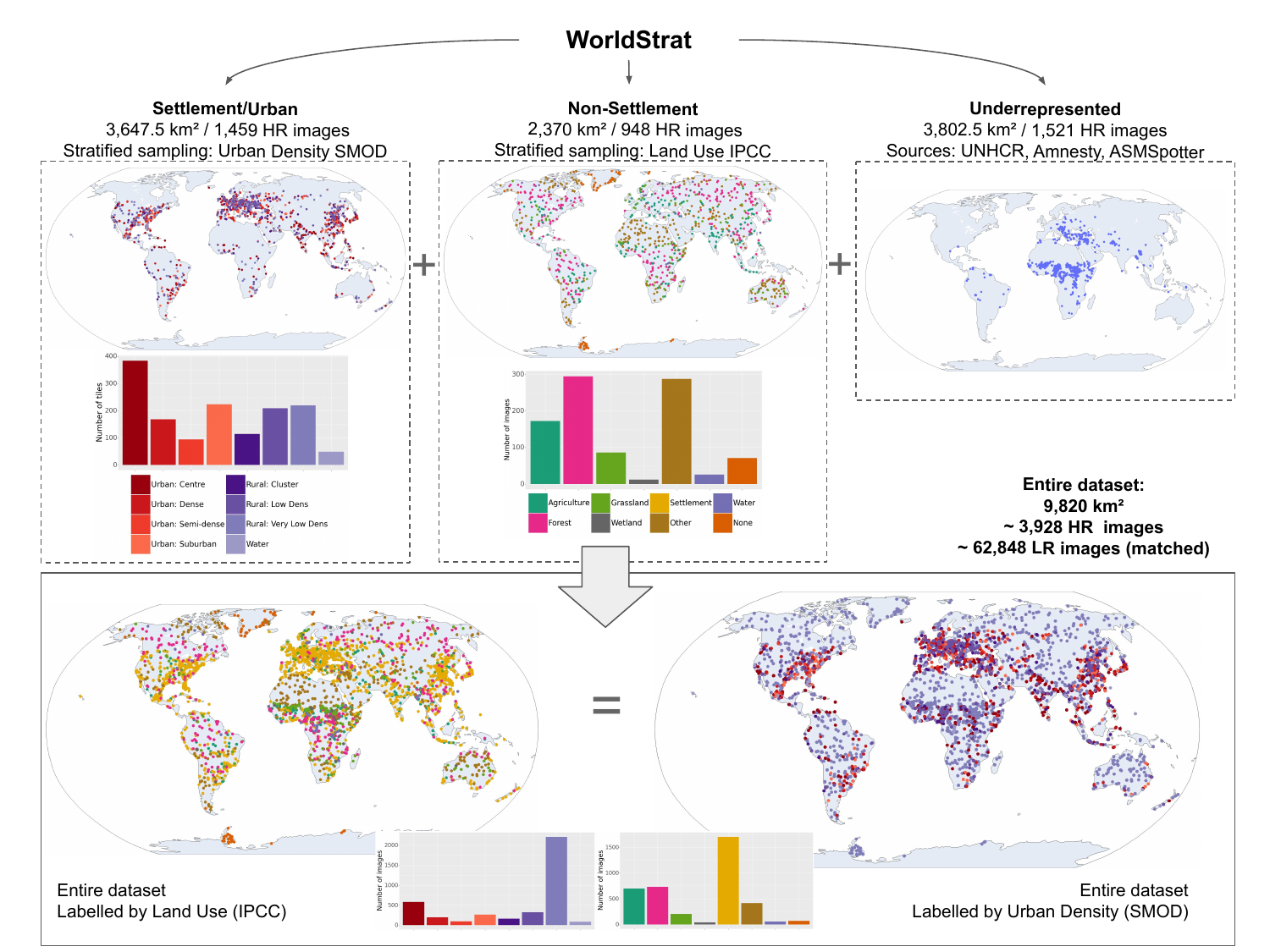}
  \caption{Summarizing the construction and classes of the WorldStrat dataset.\label{fig:summary_construction_datasheet}}
\end{figure}

We detailed our sampling procedure in 
\ifdefined\datasheetsolo
Section~2 of the accompanying article \cite{cornebise_neurips_2022},
\else
Section~\ref{sec:stratif},
\fi
with parts duplicated in the rest of this answer for the readers' convenience, including the summary in Figure~\ref{fig:summary_construction_datasheet}.
\smallskip

%\tosay{Explain this is a random sample, stratified, class imbalance, cubic root.}

We use the first half of the dataset to attempt a systematic, stratified coverage of the world.
 The question becomes: how do we chose these locations to ensure a ''best'' application-agnostic dataset for super-resolution?
 
Sixty percent was taken from the ``Settlement'' class from the ESA CCI LandCover Product, 
  which we then stratified according to the Global Human Settlement Layer
  SMOD for different types of urban density, and with marginal distribution
  proportional to the cubic root of the actual distribution -\/- to
  keep the order of classes but diminish the overall imbalance. 
  
\smallskip

 Forty  percent was taken from all the other IPCC classes, i.e.
  non-settlement,
stratified according to (non-settlement) IPCC class, marginal
  distribution proportional to the cubic root of the actual
  distribution, and within each (non-settlement) ~IPCC class, again stratifying,
  according to the LCCS class (thinner vegetation typology), again with
  cubic root proportions.
  
\smallskip

The second half of the dataset is obtained by sourcing 3,895 sq km around 1,062 Points Of Interest (POIs) from specialists of use-cases ignored by most existing datasets. For the rarer type of POIs, we sample 9 actual images in a non-overlapping grid centered on the POI.

\dsquestionex{What data does each instance consist of? “Raw” data (e.g., unprocessed text or images) or features?}{In either case, please provide a description.}

Each instance comprises of: date and time, geographical coordinates, high resolution imagery, and multiple low resolution imagery, with specifics as follows.

\textbf{Date and time} at which each satellite image was captured.
\textbf{Geographical coordinates} of the patch of land, as latitude and longitude coordinates of the center of the image, and as those of the bounding box.
   
\textbf{High Resolution imagery}: 1 image of a visit at high resolution, captured by Airbus SPOT 6/7 satellites in R. We provide both the orthorectified imagery as preprocessed by SentinelHub\footnote{\url{https://docs.sentinel-hub.com/api/latest/data/airbus/spot}}, and the raw cropped product from Airbus. The latter has black bands on the boundaries due to the lack of orthorectification. Each image has 5 channels: RGB (6 m/pixel), Near Infrared (6 m/pixel), and Pan-chromatic channels (1.5m / pixel), at 1054x1054 pixels at that highest resolution. The date of the visit has been picked at random between 2017 and 2019 amongst the visits whose whole-scene cloud-cover is lower than 5\%. Because our AOIs are much smaller than a full SPOT scene, it is not absolutely guaranteed that the actual image has precisely 5\% cloud -- it is likely to be entirely empty of clouds. This provides a good target image to reconstruct in the case of super-resolution. 

We provide two types of preprocessing high-resolution imagery: the "raw" imagery as provided by Airbus, and the orthorectified imagery as preprocessed by SentinelHub.

\textbf{Low Resolution imagery}: 16 Low-Resolution images from distinct revisits by Copernicus Sentinel 2, temporally matched to the High-Resolution image -- within 5 days for the temporally closest. All 12 spectral bands are covered, at up to 10 m/pixel.
We chose to not filter the low resolution Sentinel 2 revisits by their cloud coverage. This is to try and ensure the training distribution on the low resolution is similar to the real world use cases, where the user will want to rebuild at a given place at a given time. Algorithms should learn to ignore clouds and be able to assemble a view from the cloudless parts of the cloudy revisits.

We provide two types of preprocessing for the low-resolution imagery: the orthorectified but non-atmospherically-corrected level "L1C" (according to \citep{sentinel2_user_guide}, and level L2A which has been atmospherically corrected, i.e. the effect of the atmosphere on light has been removed according to physical models so colors look closer to the ground conditions. 
\medskip

See below in the Section "Collection" later in this datasheet for \textbf{important information about the temporal matching between low and high-resolution imagery}.

\dsquestionex{Is there a label or target associated with each instance?}{If so, please provide a description.}

\begin{figure}[htbp]
  \centering
  \includegraphics[width=.5\textwidth, height=.5\textheight, valign=t]{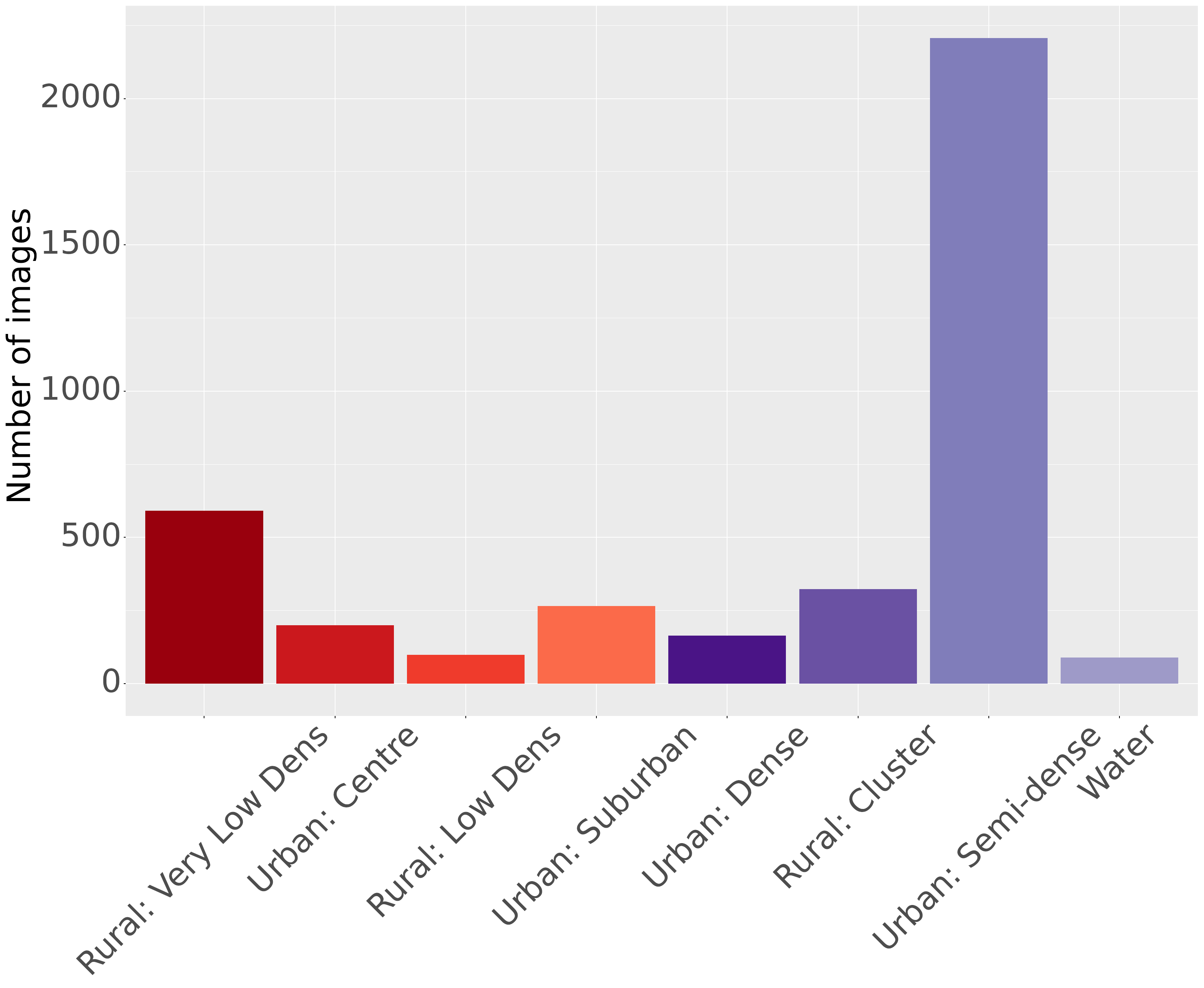}\hfill
  \includegraphics[width=.5\textwidth, height=.5\textheight,  valign=t]{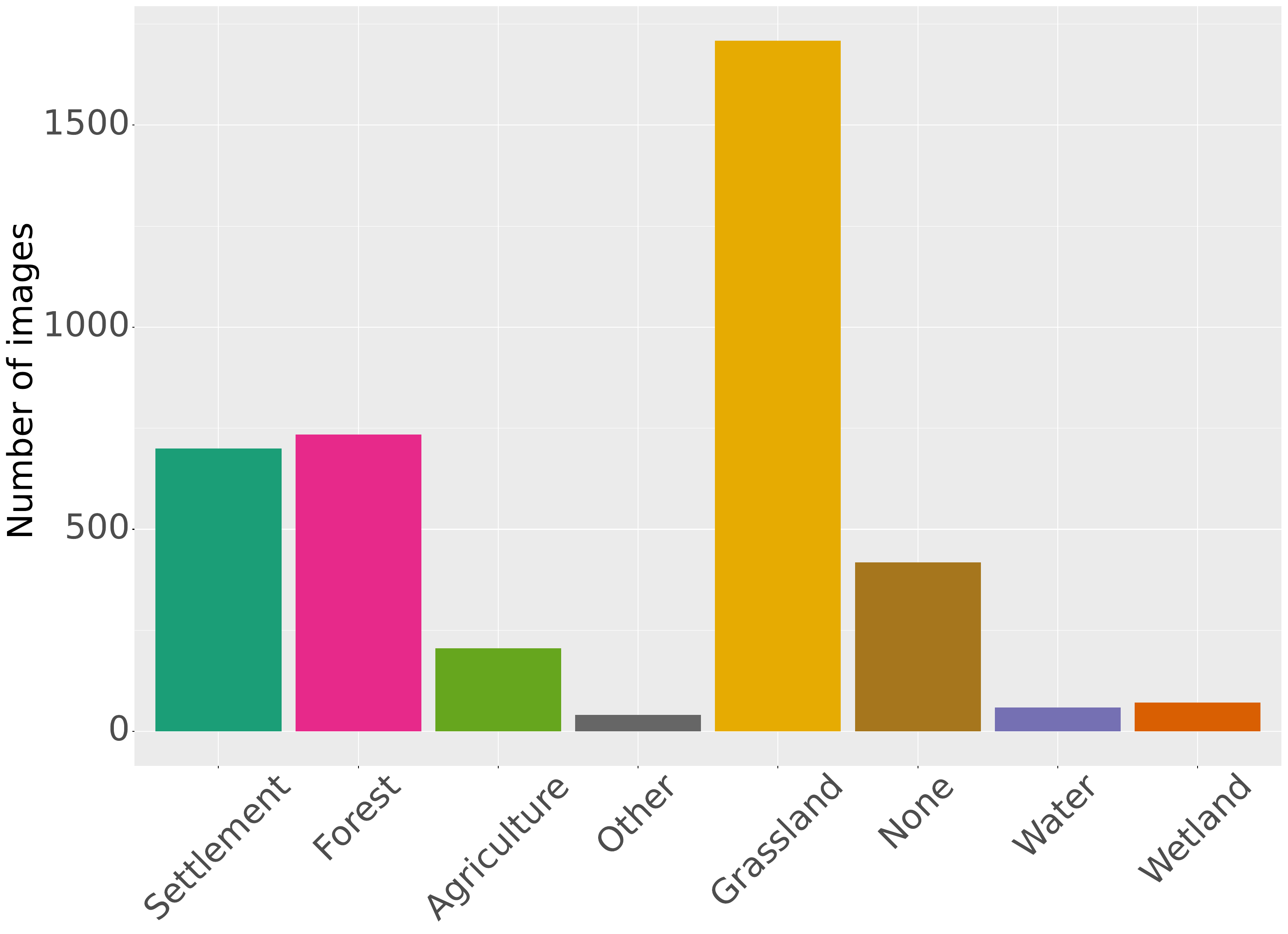}\\
  
  \includegraphics[width=1\textwidth,height=.5\textheight]{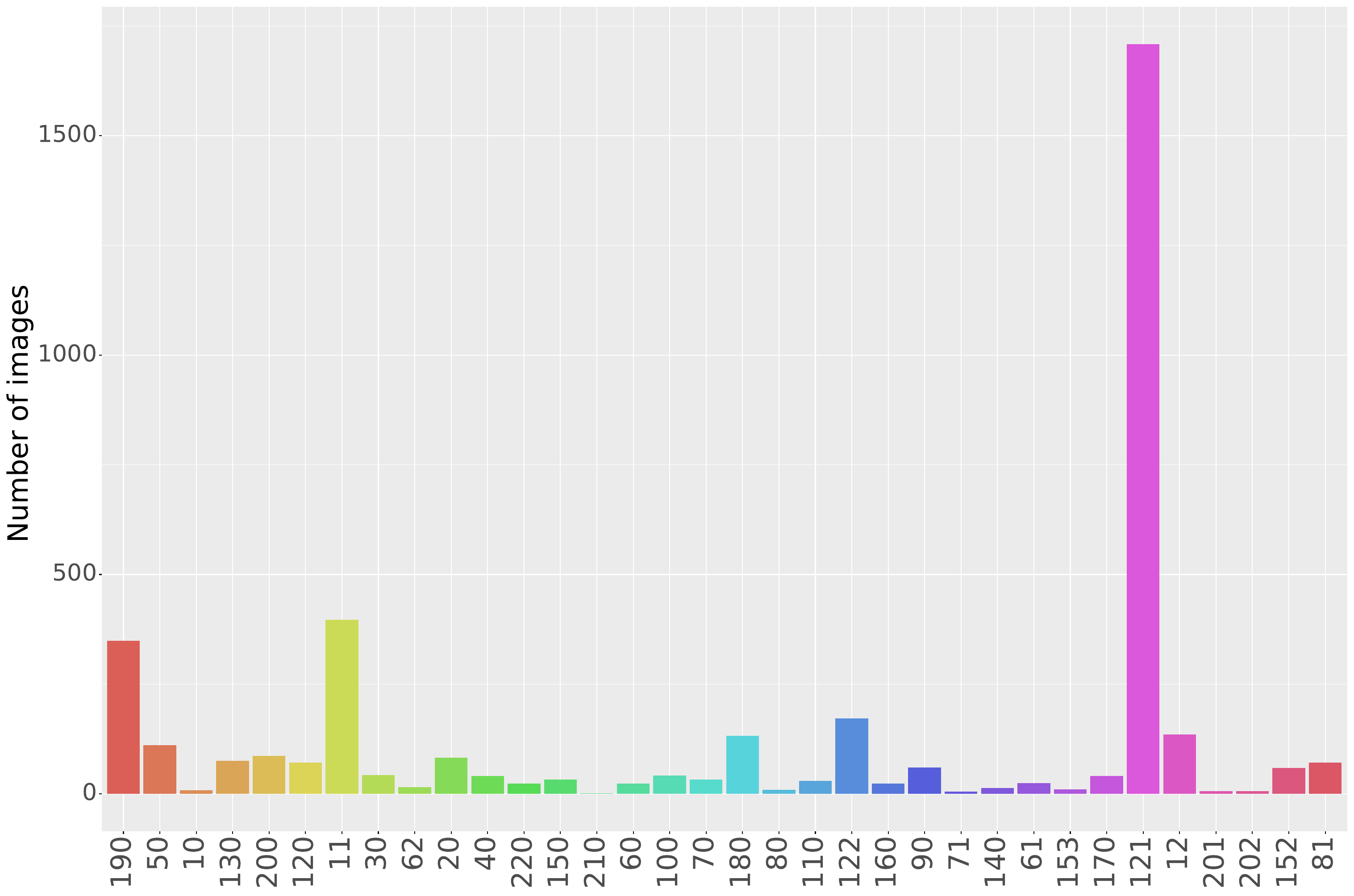}
  \caption{Frequency of class occurrences within the dataset, for SMOD urban density (top left), IPCC land use (top right), LCCS land use (bottom).\label{fig:lccs_classes_within_dataset} }
\end{figure}

The low-resolution imagery can be seen as a label for the high-resolution imagery, at least for multi-frame super-resolution tasks.
As to formal classes, we provide  three labels for each image, coming from the two datasets used to stratify the sampling (see earlier question about sampling):

\begin{itemize}
    \tightlist
    \item \textbf{Land use labels} from the European Space Agency (ESA) Climate Change Initiative (CCI) Land Cover (LC) dataset \citep{esa_land_v20}, in two forms detailed in Table~\ref{tab:esa_cci_classes}:
    \begin{itemize}
        \item The highly detailed LCCS classification, with frequencies listed in Table~\ref{tab:LCCS_frequency} and visualized in Figure~\ref{fig:lccs_classes_within_dataset} (bottom).
        \item The matching but coarser Intergovernmental Panel on Climate Change (IPCC) land categories,  with frequencies listed in Table~\ref{tab:SMOD_IPCC_frequency} (right) and visualized in Figure~\ref{fig:lccs_classes_within_dataset} (top right).
    \end{itemize}
    For more details see page 23 of \citet{esa_land_v24} and page 32 of \citet{esa_land_v20}.
    \item \textbf{Urban density label} from The Global Human Settlement Layer (GHSL) Settlement Model (SMOD) dataset, indicating the urban density, described in Table~\ref{tab:smod_table}. Class frequencies are listed in Table~\ref{tab:SMOD_IPCC_frequency} (left) and visualized in Figure~\ref{fig:lccs_classes_within_dataset} (top left). For more details see page 24 of \citet{florczyk_ghsl_2019}. 
\end{itemize}

\textbf{Important Note:} The labels correspond to the class assigned to the location at \textbf{center of the image} by the corresponding datasets. This \emph{does not mean} that the label is valid for all pixels in the image, because the spatial grids used for each of the two labeling datasets and for the images differ.

\begin{table}[htbp!]
   \centering
   \begin{tabular}{cp{9cm}cc}
   \toprule
   &                        &  Class ID \\
   IPCC Class & LCCS Class &           \\
   \midrule
   None & No Data &         0 \\
   Agriculture & Cropland, rain-fed &        10 \\
   & Herbaceous cover &        11 \\
   & Tree or shrub cover &        12 \\
   & Cropland, irrigated or post‐flooding &        20 \\
   & Mosaic cropland (\textgreater50\%) / natural vegetation (tree, shrub, herbaceous cover) (\textless50\%) &        30 \\
   & Mosaic natural vegetation (tree, shrub, herbaceous cover) (\textgreater50\%) / cropland (\textless50\%) &        40 \\
   Forest & Tree cover, broad-leaved, evergreen, closed to open (\textgreater15\%) &        50 \\
   & Tree cover, broad-leaved, deciduous, closed to open (\textgreater15\%) &        60 \\
   & Tree cover, broad-leaved, deciduous, closed (\textgreater40\%) &        61 \\
   & Tree cover, broad-leaved, deciduous, open (15‐40\%) &        62 \\
   & Tree cover, needleleaved, evergreen, closed to open (\textgreater15\%) &        70 \\
   & Tree cover, needleleaved, evergreen, closed (\textgreater40\%) &        71 \\
   & Tree cover, needleleaved, evergreen, open (15‐40\%) &        72 \\
   & Tree cover, needleleaved, deciduous, closed to open (\textgreater15\%) &        80 \\
   & Tree cover, needleleaved, deciduous, closed (\textgreater40\%) &        81 \\
   & Tree cover, needleleaved, deciduous, open (15‐40\%) &        82 \\
   & Tree cover, mixed leaf type (broad-leaved and needleleaved) &        90 \\
   & Mosaic tree and shrub (\textgreater50\%) / herbaceous cover (\textless50\%) &       100 \\
   & Tree cover, flooded, fresh or brackish water &       160 \\
   & Tree cover, flooded, saline water &       170 \\
   Grassland & Mosaic herbaceous cover (\textgreater50\%) / tree and shrub (\textless50\%) &       110 \\
   & Grassland &       130 \\
   Wetland & Shrub or herbaceous cover, flooded, fresh/saline/brackish water &       180 \\
   Settlement & Urban areas &       190 \\
   Other: Shrubland & Shrubland &       120 \\
   & Evergreen shrubland &       121 \\
   & Deciduous shrubland &       122 \\
   Other: Sparse vegetation & Lichens and mosses &       140 \\
   & Sparse vegetation (tree, shrub, herbaceous cover) (\textless15\%) &       150 \\
   & Sparse shrub (\textless15\%) &       152 \\
   & Sparse herbaceous cover (\textless15\%) &       153 \\
  
   Other: Bare area & Bare areas &       200 \\
   & Consolidated bare areas &       201 \\
   & Unconsolidated bare areas &       202 \\
   Other: Water & Water bodies &       210 \\
   None & Permanent snow and ice &       220 \\
   \bottomrule
   \end{tabular}
   \caption{Land use classes according to the LCSS and the IPCC classifications.
   LCCS comprises of 23 classes and 14 sub-classes. IPCC groups those into 6 coarser classes.\label{tab:esa_cci_classes}}
\end{table}

\begin{table}[htbp]
   \centering
   \begin{tabular}{ll}
   \toprule
   {} &              Class ID \\
   SMOD Class &                       \\
   \midrule
   30         &         Urban: Centre \\
   23         &          Urban: Dense \\
   22         &     Urban: Semi-dense \\
   21         &       Urban: Suburban \\
   13         &        Rural: cluster \\
   12         &       Rural: Low Dens \\
   11         &  Rural: Very Low Dens \\
   10         &                 Water \\
   \bottomrule
   \end{tabular}
   \caption{The GHSL-SMOD dataset comprises of 3 classes at level 1, and 8 sub-classes at level 2, as described in the GHSL Data Package.\label{tab:smod_table}}
\end{table}

\begin{table}[htbp]
   \centering
    \begin{tabular}{p{9cm}lp{1cm}l}
    \toprule
    {} &  Frequency \\
    LCCS Class                                                                         &            \\
    \midrule
    Urban areas                                                                        &       1,709 \\
    Tree cover, broad-leaved, evergreen, closed to open (\textgreater15\%)                          &        397 \\
    Cropland, rain-fed                                                                  &        349 \\
    Grassland                                                                          &        172 \\
    Bare areas                                                                         &        135 \\
    Shrubland                                                                          &        132 \\
    Herbaceous cover                                                                   &        111 \\
    Mosaic cropland (\textgreater50\%) / natural vegetation (tree, shrub, herbaceous cover) (\textless50\%) &         86 \\
    Tree cover, broad-leaved, deciduous, open (15‐40\%)                                  &         82 \\
    Cropland, irrigated or post‐flooding                                               &         75 \\
    Mosaic natural vegetation (tree, shrub, herbaceous cover) (\textgreater50\%) / cropland (\textless50\%) &         71 \\
    Permanent snow and ice                                                             &         71 \\
    Sparse vegetation (tree, shrub, herbaceous cover) (\textless15\%)                           &         60 \\
    Water bodies                                                                       &         59 \\
    Tree cover, broad-leaved, deciduous, closed to open (\textgreater15\%)                          &         43 \\
    Mosaic tree and shrub (\textgreater50\%) / herbaceous cover (\textless50\%)                             &         42 \\
    Tree cover, needleleaved, evergreen, closed to open (\textgreater15\%)                         &         41 \\
    Shrub or herbaceous cover, flooded, fresh/saline/brakish water                     &         41 \\
    Tree cover, needleleaved, deciduous, closed to open (\textgreater15\%)                         &         33 \\
    Mosaic herbaceous cover (\textgreater50\%) / tree and shrub (\textless50\%)                             &         33 \\
    Deciduous shrubland                                                                &         29 \\
    Tree cover, flooded, fresh or brackish water                                        &         24 \\
    Tree cover, mixed leaf type (broad-leaved and needleleaved)                         &         23 \\
    Tree cover, needleleaved, evergreen, closed (\textgreater40\%)                                 &         23 \\
    Lichens and mosses                                                                 &         23 \\
    Tree cover, broad-leaved, deciduous, closed (\textgreater40\%)                                  &         15 \\
    Sparse herbaceous cover (\textless15\%)                                                     &         13 \\
    Tree cover, flooded, saline water                                                  &         10 \\
    Evergreen shrubland                                                                &          9 \\
    Tree or shrub cover                                                                &          8 \\
    Consolidated bare areas                                                            &          6 \\
    Unconsolidated bare areas                                                          &          6 \\
    Sparse shrub (\textless15\%)                                                                &          5 \\
    Tree cover, needleleaved, deciduous, closed (\textgreater40\%)                                 &          1 \\
    \bottomrule
    \end{tabular}
    \caption{Frequency of LCCS class occurrences in the dataset.\label{tab:LCCS_frequency}}
\end{table}

\begin{table}[htbp]
   \centering
    \begin{tabular}{ccc}
    \toprule
    {} &  Frequency \\
    SMOD Class           &            \\
    \midrule
    Rural: Very Low Dens &       2,207 \\
    Urban: Centre        &        591 \\
    Rural: Low Dens      &        323 \\
    Urban: Suburban      &        265 \\
    Urban: Dense         &        200 \\
    Rural: Cluster       &        164 \\
    Urban: Semi-dense    &         98 \\
    Water                &         89 \\
    \bottomrule
    \end{tabular}
    \hfill
    \begin{tabular}{ccc}
    \toprule
    {} &  Frequency \\
    IPCC Class  &            \\
    \midrule
    Settlement  &       1,709 \\
    Forest      &        734 \\
    Agriculture &        700 \\
    Other       &        418 \\
    Grassland   &        205 \\
    None        &         71 \\
    Water       &         59 \\
    Wetland     &         41 \\
    \bottomrule
    \end{tabular}
    \caption{Frequency of SMOD (left) and IPCC (right) class occurrences in the dataset. \label{tab:SMOD_IPCC_frequency}}
\end{table}

\dsquestionex{Is any information missing from individual instances?}{If so, please provide a description, explaining why this information is missing (e.g., because it was unavailable). This does not include intentionally removed information, but might include, e.g., redacted text.}

Optical satellite observation, always risks suffering from obstruction due to cloud coverage. We have selected the high-resolution images for the lowest cloud-covering at the "scene" level, i.e. the larger product area containing which the 2.5\sqkm tile. This lowers considerably the probability of cloud on the small tile we purchased, but does not exclude it entirely. Of course this biases the visit dates towards sunny seasons. We did not do any such filtering on the lower-resolution imaggery, to represent typical sampling conditions.

\dsquestionex{Are relationships between individual instances made explicit (e.g., users’ movie ratings, social network links)?}{If so, please describe how these relationships are made explicit.}

Any spatial relationship between the instances is given by the geolocation data provided with each tile. Note that, to the best of our verifications, there should be no overlaps between tiles.

\dsquestionex{Are there recommended data splits (e.g., training, development/validation, testing)?}{If so, please provide a description of these splits, explaining the rationale behind them.}

We do provide a recommended split, between training, validation, and testing, for easy comparison. We have used approximately 80\% / 10 \% / 10 \% proportions amongst the splits, by stratified random sampling amongst the joint classes, deffined as the triplets (source class, IPCC class, SMOD class). Joint classes with only one AOI (15 AOIs, i.e. 0.3856\% of the dataset) were all assigned to the training set, and those for which there was only one AOI left *after* assignment to the training set were randomly assigned to either testing or validation set, uniformly (12 AOIs).

\begin{figure}[htbp]
  \centering
  \includegraphics[height=1\textwidth]{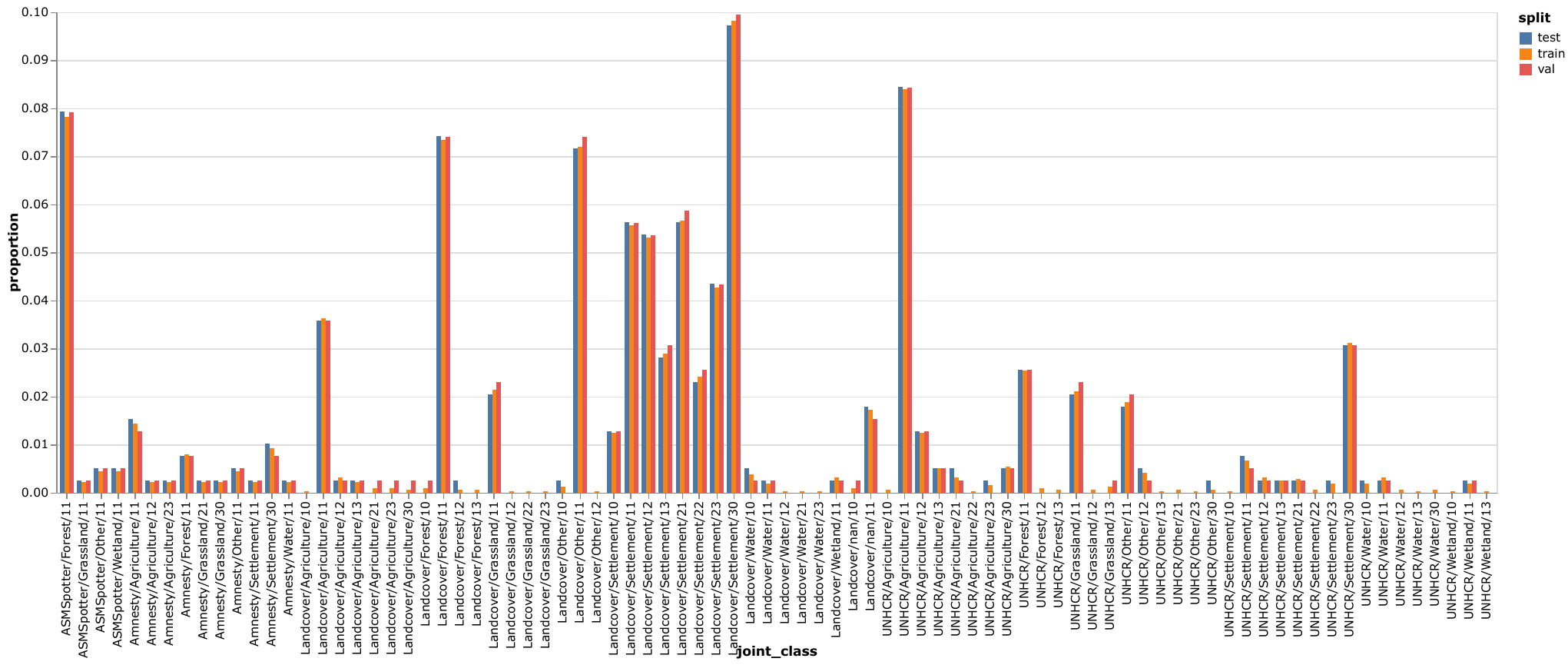}
  \caption{Proportions of all joint classes in the training, testing, and validation sets in the recommended data split provided as part of the dataset. Stratification ensures minimum differences between the proportions amongst the three splits.\label{fig:split_histogram}}
\end{figure}

Note: v1.0 of the dataset came with a different and erroneous split (see Changelog at the beginning of this datasheet). For this reason we recommend ensuring that you use v1.1 or above of the dataset. If the MD5 sum of file \texttt{stratified\char`_train\char`_val\char`_test\char`_split.csv} is \texttt{745035835d835280aa0298a9dc1996d1}, you are still using v1.0 and should update to a new version of the dataset.

\dsquestionex{Are there any errors, sources of noise, or redundancies in the dataset?}{If so, please provide a description.}

For the under-represented areas, UNHCR locations are not guaranteed to be locately absolutely precisely on settlements of persons at risk. This is somewhat mitigated by the area of our tiles. In addition, in case they are not, it is important to know that such settlements are often similar to neighbouring constructions: the visual features on the tile can therefore be considered as representative. 

As discussed in the previous questions, the spatial grids used for each of the two labeling datasets and for the images have different origins and different resolutions: therefore, all we can guarantee is that the label applies to the corresponding dataset's tile containing central pixel of the imagery.

As also explained in the question on pre-processing, there is a redundancy between the orthorectified and the non-orthorectified high-resolution imagery, and between the L1C and L2A (atmospherically corrected) low-resolution imagery.

\dsquestionex{Is the dataset self-contained, or does it link to or otherwise rely on external resources (e.g., websites, tweets, other datasets)?}{If it links to or relies on external resources, a) are there guarantees that they will exist, and remain constant, over time; b) are there official archival versions of the complete dataset (i.e., including the external resources as they existed at the time the dataset was created); c) are there any restrictions (e.g., licenses, fees) associated with any of the external resources that might apply to a future user? Please provide descriptions of all external resources and any restrictions associated with them, as well as links or other access points, as appropriate.}

The dataset is self-contained.

For ease of use, however, we have included the class information from the ESA CCI LC dataset\citep{esa_land_v20}, the UNHCR Persons of Concern dataset \citep{unhcr_2021}, the images from Copernicus Sentinel 2 archive \citep{drusch2012sentinel}, along with the Airbus SPOT\citep{spot_user_guide} imagery that we have acquired.
None of those have a link to timestamped versions as far as we are aware, hence the inclusion of the subset used in this dataset is more than mere convenience, it is also archiving for consistency.

The license on the dataset has been carefully chosen to be compatible with the licenses of each of these original sources: we release the labels and the low-resolution imagery under Creative Commons with Attribution 4.0 International (CC BY 4.0\footnote{\url{https://creativecommons.org/licenses/by/4.0/}}, and the high-resolution imagery from Airbus is distributed, with authorization from Airbus, under Creative Commons Attribution-NonCommercial 4.0 International (CC BY-NC 4.0)\footnote{\url{https://creativecommons.org/licenses/by-nc/4.0/}}.

We have used SentinelHub (\url{https://www.sentinel-hub.com/}) to download the satellite imagery used in this dataset. All the code used for the construction of the dataset is included in the accompanying Python package.

\dsquestionex{Does the dataset contain data that might be considered confidential (e.g., data that is protected by legal privilege or by doctor-patient confidentiality, data that includes the content of individuals non-public communications)?}{If so, please provide a description.}

There is no confidential data in this dataset.

\dsquestionex{Does the dataset contain data that, if viewed directly, might be offensive, insulting, threatening, or might otherwise cause anxiety?}{If so, please describe why.}

This dataset contains satellite imagery of sites classified by UNHCR as hosting Persons of Concern: vulnerable populations such as displaced populations, refugee camps, locations close to conflict zones. It also contains sites of interest for Human Rights investigations, such as prisons or jails.

\dsquestionex{Does the dataset relate to people?}{If not, you may skip the remaining questions in this section.}

Yes, as there is human density information and inclusion of locations hosting Persons of Concern, but only in aggregate. These locations were already public in the UNHCR Persons of Concerns dataset.

\dsquestionex{Does the dataset identify any subpopulations (e.g., by age, gender)?}{If so, please describe how these subpopulations are identified and provide a description of their respective distributions within the dataset.}

The dataset contains locations and satellite imagery of locations hosting "Persons of Concerns" as classified by UNHCR. This represents 981\sqkm out of a total of 9,820\sqkm.

\dsquestionex{Is it possible to identify individuals (i.e., one or more natural persons), either directly or indirectly (i.e., in combination with other data) from the dataset?}{If so, please describe how.}

No.

\dsquestionex{Does the dataset contain data that might be considered sensitive in any way (e.g., data that reveals racial or ethnic origins, sexual orientations, religious beliefs, political opinions or union memberships, or locations; financial or health data; biometric or genetic data; forms of government identification, such as social security numbers; criminal history)?}{If so, please provide a description.}

The location of refugee populations such as listed in UNHCR "Persons of Concerns" dataset and included in this dataset might be considered sensitive. UNHCR already published these locations, and is better placed than us to decide on whether this was sensitive. The locations of some sites of interest for Human Rights investigation, however, are newly listed. And of course, this is the first time that satellite imagery of all these locations is published in one single dataset, to the best of our knowledge -- although these are also accessible on common mapping websites for the casual viewer.

See question on ethic reviews later in this datasheet for more details.

\dsquestion{Any other comments?}

None.

%%%%%%%%%%%%%%%%%%%%%%%%%%%%%%%%%%%%%%%%%%
\bigskip
\dssectionheader{Collection Process}

\dsquestionex{How was the data associated with each instance acquired?}{Was the data directly observable (e.g., raw text, movie ratings), reported by subjects (e.g., survey responses), or indirectly inferred/derived from other data (e.g., part-of-speech tags, model-based guesses for age or language)? If data was reported by subjects or indirectly inferred/derived from other data, was the data validated/verified? If so, please describe how.}

As described above, the imagery was acquired using Airbus SPOT 6/7 and Copernicus Sentinel 2 satellites, downloaded with SentinelHub. We refer to their user guides \citep{spot_user_guide}, as well as to SentinelHub FAQ\footnote{\url{https://www.sentinel-hub.com/faq/}} for details of their internal processing.

% SentinelHub User Guide (Has S2 L2A/L1C and SPOT): https://docs.sentinel-hub.com/api/latest/data/
% SentinelHub FAQ: https://www.sentinel-hub.com/faq/
% Airbus SPOT 6/7 User Guide: https://www.intelligence-airbusds.com/#spot6

\dsquestionex{What mechanisms or procedures were used to collect the data (e.g., hardware apparatus or sensor, manual human curation, software program, software API)?}{How were these mechanisms or procedures validated?}

We open-source software for the collection of the data in the accompanying Python package, done using parsing of the CSV of the land uses datasets, sampling of locations using pseudo-random number generators and stratified sampling, and SentinelHub API for ordering and downloading the satellite imagery.

\dsquestion{If the dataset is a sample from a larger set, what was the sampling strategy (e.g., deterministic, probabilistic with specific sampling probabilities)?}

We refer to the "Composition" section of this datasheet for the description of our sampling methodology.

\dsquestion{Who was involved in the data collection process (e.g., students, crowdworkers, contractors) and how were they compensated (e.g., how much were crowdworkers paid)?}

The data was assembled by the persons cited as the author of this dataset in the first question. 
The actual collection of the satellite imagery was done by Airbus and Sentinel 2 / ESA, and we refer to the documentation of the UNHCR, ESA CCI, ASMSpotter, and GHSL SMOD for their respective collection methodologies. The Airbus imagery was acquired as part of ESA Third Party Mission program (TPM) and the license extension for distribution was negotiated and paid for specially by ESA TPM.

\dsquestionex{Over what timeframe was the data collected? Does this timeframe match the creation timeframe of the data associated with the instances (e.g., recent crawl of old news articles)?}{If not, please describe the timeframe in which the data associated with the instances was created.}

\begin{itemize}
    \tightlist
    \item Actual assembly of the dataset took place from March 2021 to June 2022.
    \item The satellite imagery was filtered to have been taken between 2017 and 2021. 
    \item Each satellite image comes with the timestamp of its acquisition.
    \item We could not find temporal information about the date of collection by UNHCR of its dataset of locations of "Persons of Concerns". We downloaded thesed locations in September 2021.
    \item We used the 2020 ESA CCI LC map as base level.  
    \item We used the R2019A release of the GHSL SMOD data \citep{pesaresi2019ghs} -- see its Product User Guide \citep{florczyk_ghsl_2019}, sections "Input Data", page 25.
    
\end{itemize}

\textbf{Matching Low-Resolution and High-Resolution Imagery}:

Unlike the low-resolution Sentinel2, the high-resolution SPOT is only available where and when it has been tasked. This raises the question of how what date to we pick for the SPOT high-resolution imagery at a location, if multiple are available, and how do we temporally match the Sentinel2 imagery, within which time window around the date of the SPOT visit? 

In theory we could try to pick POIs and SPOT visit times that maximise the number of Sentinel2 imagery available within a fixed length time window, so as to have the richest training set.  We do indeed observe some variation in that regard: some of these lines go much higher than others.

However, this would induce an implicit bias of a nature hard to interpret. We also observe that within a POI, the discrepancy between the number of S2 revisits, while clearly present, is reasonable, with multiple SPOT revisits offering similar S2 availability. 

Finally, biasing per Sentinel2 imagery would be akin to biasing per Sentinel2 cloud coverage: this would not be a fair representation of real-world use cases, and we would therefore be training our models for the wrong problem.

We therefore took the decision yet again to solve a harder problem than an optimally-curated dataset would make for, so as to be the closest to reality. To that effect, within a POI, we pick uniformly at random the SPOT visit to use as a reference.

Of course, one bias remains: we will not have imagery of POIs that have never been tasked by SPOT customers. While unfortunate, there is no way around it, short of using another high resolution product. We do have hope in two mitigating factors:
\begin{itemize}
\tightlist
  \item SPOT swath covers more than just the single POI, so we cover areas that are possibly more diverse than just the one precise point of interest to the SPOT customer.
  \item SPOT tasking means the POI exhibits features of activity interesting to at least the SPOT customer. SPOT customers might not have entirely the same interests as the users of our open-source package, but it is not unreasonable to assume that the features will be transferable. Therefore, this implicit sampling is actually a positive way to ensure interesting features.
\end{itemize}

\dsquestionex{Were any ethical review processes conducted (e.g., by an institutional review board)?}{If so, please provide a description of these review processes, including the outcomes, as well as a link or other access point to any supporting documentation.}

No IRB was required. However, we wanted to be careful due to the sensitivity of UNHCR Persons of Concerns locations and the locations provided by Amnesty International. It is important to note that the UNHCR dataset of locations was already published -- but not with satellite imagery collated this way.

We therefore consulted with other human rights experts not otherwise involved in this project before releasing this dataset. They pointed at the precedent of, for example, Human Rights Watch (HRW) releasing the map of torture sites in Syria in 2012\footnote{\url{https://www.hrw.org/video-photos/interactive/2012/07/02/interactive-map-syrias-torture-centers}}$^{,}$\footnote{\url{https://www.hrw.org/news/2012/07/03/syria-torture-centers-revealed}}. In that particular case, emphasis was put by the persons on the ground, living near those locations, that these sites were already widely known to local forces. Publication of these locations therefore provided limited extra risk by local exposure, and provided large benefits from the worldwide attention attracted by their global exposure.

Based on these precedents, and, most importantly, on the fact that all this information was already available albeit not in this packaged form, we decided to publish.

\dsquestionex{Does the dataset relate to people?}{If not, you may skip the remaining questions in this section.}

Only to people in aggregate populations, via SMOD urban density and UNHCR locations.

\dsquestion{Did you collect the data from the individuals in question directly, or obtain it via third parties or other sources (e.g., websites)?}

As explained above, we obtained that data about populations via GHSL SMOD and UNHCR.

\dsquestionex{Were the individuals in question notified about the data collection?}{If so, please describe (or show with screenshots or other information) how notice was provided, and provide a link or other access point to, or otherwise reproduce, the exact language of the notification itself.}

We do not have access to this information for GHSL SMOD and UNHCR data products. The construction of GHSL SMOD involved in part administrative data such as census, so answering this question accurately would need tracing each individual census they obtained. 

We reinforce that we are probably over-cautious in answering this question, as we take a very broad view of the term "relate to people" in the question. These are population-wide density data, and locations of settlements, already published.

\dsquestionex{Did the individuals in question consent to the collection and use of their data?}{If so, please describe (or show with screenshots or other information) how consent was requested and provided, and provide a link or other access point to, or otherwise reproduce, the exact language to which the individuals consented.}

See above.

\dsquestionex{If consent was obtained, were the consenting individuals provided with a mechanism to revoke their consent in the future or for certain uses?}{If so, please provide a description, as well as a link or other access point to the mechanism (if appropriate).}

Not applicable

\dsquestionex{Has an analysis of the potential impact of the dataset and its use on data subjects (e.g., a data protection impact analysis) been conducted?}{If so, please provide a description of this analysis, including the outcomes, as well as a link or other access point to any supporting documentation.}

See above about ethical discussion of impact.

\dsquestion{Any other comments?}

None.

%%%%%%%%%%%%%%%%%%%%%%%%%%%%%%%%%%%%%%%%%%
\bigskip
\dssectionheader{Preprocessing/cleaning/labeling}

\dsquestionex{Was any preprocessing/cleaning/labeling of the data done (e.g., discretization or bucketing, tokenization, part-of-speech tagging, SIFT feature extraction, removal of instances, processing of missing values)?}{If so, please provide a description. If not, you may skip the remainder of the questions in this section.}

The preprocessing of the imagery was explained above as part of the collection -- in remote sensing, the separation between collection and preprocessing is very arbitrary, as they form a continuous spectrum of more and more refined data products.

\dsquestionex{Was the “raw” data saved in addition to the preprocessed/cleaned/labeled data (e.g., to support unanticipated future uses)?}{If so, please provide a link or other access point to the “raw” data.}

As described above, we provide both "raw" and orthorectified Airbus data, and L1C and L2A Sentinel 2 imagery.

\dsquestionex{Is the software used to preprocess/clean/label the instances available?}{If so, please provide a link or other access point.}

Yes. We provide in the accompanying Python package the entire code we used to collect, sample, assemble, and pre-process the data.

\dsquestion{Any other comments?}

The processing of satellite imagery is a very complex topic with a long history and a huge domain expertise required - few people master it end-to-end, certainly not us. We wanted to lower the barrier to entry by providing this dataset, and the accompanying PyTorch DataLoader, in a format most accessible to the Machine Learning community. Remote sensing experts might therefore frown upon the levity with which we (do not) discuss many details of satellite imagery (e.g. angle of incidence, atmospheric collection models, etc). We have provided throughout this datasheet the references for anyone interested in tracing all the processing steps by the providers of the individual data products.

%%%%%%%%%%%%%%%%%%%%%%%%%%%%%%%%%%%%%%%%%%
\bigskip
\dssectionheader{Uses}

\dsquestionex{Has the dataset been used for any tasks already?}{If so, please provide a description.}

Yes. We have used this dataset for a benchmark comparison of three baseline multi-frame super-resolution algorithm, aiming at super-resolving from 10m/pixel (Sentinel 2) multi-spectral to 3m/pixel obtained by down-sampling the Pan-Sharpened SPOT 6/7 imagery. We refer to 
\ifdefined\datasheetsolo
Section~4 of 
the accompanying article~\citep{cornebise_neurips_2022},
\else
Section~\ref{sec:using} of the main article
\fi
and to the accompanying Python package which allows full reproduction of that benchmark. These are meant as baseline to illustrate how to use this dataset, and we have entire confidence that they will be beat very soon -- we are looking forward to users training their own algorithms!

\dsquestionex{Is there a repository that links to any or all papers or systems that use the dataset?}{If so, please provide a link or other access point.}

This dataset is archived on Zenodo with the DOI \url{https://doi.org/10.5281/zenodo.6810791}, which allows to search for all papers citing it in any bibliographic database such as Google Scholar. Zenodo also maintains automatically a list of uses and citations to the dataset. It does not allow to manually add uses that do not cite the DOI e.g. because they were not accompanied by a publication. It does not either track the citations to the NeurIPS article where we present the dataset. We do not currently have plans to manually maintain a separate repository of usages, but could be convinced to do so if several users request it.

\dsquestion{What (other) tasks could the dataset be used for?}

As explained in the Introduction of the accompanying article, we have designed this dataset so it can be used for the broadest range of machine learning applications for satellite imagery. 

Of course, one immediate use is to further super-resolution research. We believe efficient super-resolution algorithms, in particular from Sentinel 2, can unlock use cases where high-resolution is not available, either due to cost or limited tasking or simply scale -- Sentinel 2 being a remarkable resource re-visiting the world every 5 days, accessible to everyone. Our benchmark 

Beyond that, we do not pretend to substitute our imagination for the creativity of our colleagues in the community. With the imagery alone, we can imagine any kind of computer vision tasks involving self-supervised or un-supervised representations on low and high resolutions, transfer tasks from one resolution to the other. We could imagine some classification tasks with the labels, with the caveats mentioned earlier on how these were used as a guideline for a rich sampling and might have temporal mismatch. Because every image is geo-referenced and timestamped, it is also possible to cross-reference it with any other source of label, for example mapping databases like OpenStreetMap, for building imprints, structure detection, etc.

By providing the code we used to create the dataset, we also make it very easy to extend its sampling using the same procedure, to obtain imagery new locations, by anyone having access to different high-resolution imagery -- Sentinel 2 low-resolution imagery being accessible to everyone already. Whether it be to redistribute or for their own use is up to the user.

\dsquestionex{Is there anything about the composition of the dataset or the way it was collected and preprocessed/cleaned/labeled that might impact future uses?}{For example, is there anything that a future user might need to know to avoid uses that could result in unfair treatment of individuals or groups (e.g., stereotyping, quality of service issues) or other undesirable harms (e.g., financial harms, legal risks) If so, please provide a description. Is there anything a future user could do to mitigate these undesirable harms?}

We listed throughout this datasheet (and mentioned again in the last answer) several limitations, in particular the temporal matching of the labels with the imagery. That limitation does not impede the variance-reduction and representativity of the dataset, but it does add noise for e.g. classification tasks. 

Other than that, there are no impact on future uses that we can think of -- and the ability for the user to easily extend the dataset if they get access to new imagery should help ensure its longevity. 

\dsquestionex{Are there tasks for which the dataset should not be used?}{If so, please provide a description.}

   None to the best of our knowledge.

\dsquestion{Any other comments?}

None.

%%%%%%%%%%%%%%%%%%%%%%%%%%%%%%%%%%%%%%%%%%
\bigskip
\dssectionheader{Distribution}

\dsquestionex{Will the dataset be distributed to third parties outside of the entity (e.g., company, institution, organization) on behalf of which the dataset was created?}{If so, please provide a description.}

Yes, this dataset is made open access as it was designed for the broadest use.
We purposefully chose the least restrictive licenses allowed for this dataset to foster reuse and hopefully upstream contributions. Only the high-resolution imagery has a restriction to non-commercial uses, as a requirement from the imagery provider Airbus.

\dsquestionex{How will the dataset will be distributed (e.g., tarball on website, API, GitHub)}{Does the dataset have a digital object identifier (DOI)?}

The dataset is distributed via Zenodo, a CERN-backed repository for datasets and code, which also provides a DOI for the dataset and a separate DOI for each updated version. Zenodo also takes care of all meta-data formatting for easy discovery.

The DOI of the dataset is: \url{https://doi.org/10.5281/zenodo.6810791}.

The accompanying Python package is distributed on Github, and packaged for the popular Python packaging managers (PyPI, Conda, etc). 
As part of the package, we also distribute several tutorials in the form of Jupyter notebooks.

\dsquestion{When will the dataset be distributed?}

The dataset is distributed on Zenodo as of July 13th, 2022.

\dsquestionex{Will the dataset be distributed under a copyright or other intellectual property (IP) license, and/or under applicable terms of use (ToU)?}{If so, please describe this license and/or ToU, and provide a link or other access point to, or otherwise reproduce, any relevant licensing terms or ToU, as well as any fees associated with these restrictions.}

As discussed above, we have worked hard to ensure the broadest diffusion possible, including on the licensing front:
\begin{itemize}
\tightlist
\item The high-resolution Airbus imagery is distributed, with authorization from Airbus, under Creative Commons Attribution-NonCommercial 4.0 International (CC BY-NC 4.0)\footnote{\url{https://creativecommons.org/licenses/by-nc/4.0/}}.
\item The labels, Sentinel2 imagery, and trained weights are released under Creative Commons with Attribution 4.0 International (CC BY 4.0\footnote{\url{https://creativecommons.org/licenses/by/4.0/}}.
\item The source code under 3-Clause BSD license\footnote{\url{https://opensource.org/licenses/BSD-3-Clause}}.
\end{itemize}

\dsquestionex{Have any third parties imposed IP-based or other restrictions on the data associated with the instances?}{If so, please describe these restrictions, and provide a link or other access point to, or otherwise reproduce, any relevant licensing terms, as well as any fees associated with these restrictions.}

As explained above, while, thanks to ESA Phi-Lab and ESA Third Party Missions, we secured license from Airbus to distribute the high-resolution imagery, that specific part of the dataset is be used only for non-commercial purposes according to the terms of the CC-BY-NC license.

\dsquestionex{Do any export controls or other regulatory restrictions apply to the dataset or to individual instances?}{If so, please describe these restrictions, and provide a link or other access point to, or otherwise reproduce, any supporting documentation.}

   None.

\dsquestion{Any other comments?}

   None.

%%%%%%%%%%%%%%%%%%%%%%%%%%%%%%%%%%%%%%%%%%
\bigskip
\dssectionheader{Maintenance}

\dsquestion{Who will be supporting/hosting/maintaining the dataset?}

Long-term maintenance of the content of the dataset will be by the authors, like every academic.

In terms of hosting, to ensure the maximum availability and long-term life, the dataset is hosted on Zenodo, which is backed by CERN. This is becoming the gold standard in terms of dataset distribution, and provides more than reasonable availability and redundancy. The underlying Zenodo infrastructure and redundancy measures are documented at Zenodo's About - Infrastructure page\footnote{\url{https://about.zenodo.org/infrastructure/}}. 

\dsquestion{How can the owner/curator/manager of the dataset be contacted (e.g., email address)?}

Julien Cornebise can be contacted at his email address at University College London: \url{j.cornebise@ucl.ac.uk}. In case of any future change of affiliation, he can also be reached at the stable address \url{julien@cornebise.com}.

\dsquestionex{Is there an erratum?}{If so, please provide a link or other access point.}

An ongoing list of errors and fixes is available in the Issues section of the Github repository of the dataset at \url{https://github.com/worldstrat/worldstrat/issues}.

\dsquestionex{Will the dataset be updated (e.g., to correct labeling errors, add new instances, delete instances)?}{If so, please describe how often, by whom, and how updates will be communicated to users (e.g., mailing list, GitHub)?}

We will be uploading any modification of the dataset to Zenodo, which will provide a version-specific DOI along with the root fixed DOI \url{https://doi.org/10.5281/zenodo.6810791} covering the ensemble of versions. We do plan to correct errors that we are made aware of, and we welcome contributions of extra imagery!

Zenodo does not yet seem to have a subscription mechanism to automatically notify subscribers of extra information, therefore we will likely post updates on GitHub or set up a mailing list -- but this is still to be determined.

\dsquestionex{If the dataset relates to people, are there applicable limits on the retention of the data associated with the instances (e.g., were individuals in question told that their data would be retained for a fixed period of time and then deleted)?}{If so, please describe these limits and explain how they will be enforced.}

   No retention limits.

\dsquestionex{Will older versions of the dataset continue to be supported/hosted/maintained?}{If so, please describe how. If not, please describe how its obsolescence will be communicated to users.}

Since the dataset is hosted on Zenodo, and Zenodo supports DOI versioning, all the different versions of the dataset are tracked and hosted. The DOI versioning functions similarly to an incremental update, duplicating only the modified files. More details can be found on Zenodo's FAQ, under DOI versioning. \footnote{\url{https://help.zenodo.org/\#versioning}}

\dsquestionex{If others want to extend/augment/build on/contribute to the dataset, is there a mechanism for them to do so?}{If so, please provide a description. Will these contributions be validated/verified? If so, please describe how. If not, why not? Is there a process for communicating/distributing these contributions to other users? If so, please provide a description.}

We welcome contributions by any generous users. They are welcome to contact us (see above) to discuss, and we will verify and validate on a case-by-case basis, as well as publish any extra code that will have been used for the enrichment. As mentioned above, we provide the source code we used to build the dataset, which makes it easy for any would-be contributor to help ensurin0g similar sampling distribution and formatting. Please do contact us ahead of time, we will be delighted to discuss how to enrich this community resource!

\dsquestion{Any other comments?}

None.

\clearpage
\listoffigures 
\addcontentsline{toc}{section}{List of Figures}
\listoftables 
\addcontentsline{toc}{section}{List of Tables}

\end{document}